\documentclass[twocolumn,showpacs,preprintnumbers,amsmath,amssymb,prb]{revtex4}

\usepackage{graphicx}% Include figure files
\usepackage{dcolumn}% Align table columns on decimal point
\usepackage{epsf}% bold math

\begin{document}

\title{Nonlinear dynamics of vortices in easy flow channels along grain boundaries in superconductors}
\author{A. Gurevich} 
\affiliation{Applied Superconductivity Center, 
University of Wisconsin, Madison, Wisconsin} 

\date{\today}

\begin{abstract}
A theory of nonlinear dynamics of mixed Abrikosov vortices with Josephson cores 
(AJ vortices) on low-angle grain boundaries (GB) in superconductors is proposed. 
As the misorientation angle $\vartheta$ increases, vortices on low-angle GBs evolve from 
the Abrikosov vortices with normal cores to intermediate AJ vortices with Josephson 
cores, whose length $l$ along GB is smaller that the London penetration depth $\lambda$, but 
larger than the coherence length $\xi$. Dynamics and pinning of the AJ vortex structures determine the in-field 
current transport through GB and the microwave response of polycrystal in the crucial 
misorientation range $\vartheta < 20-30^{\circ}$ of the exponential drop of the local critical 
current density $J_b(\vartheta)$ through GB. An exact solution for an overdamped periodic AJ vortex 
structure driven along GB by an arbitrary time dependent transport current in a dc magnetic field 
$H>H_{c1}$ is obtained. It is shown that the dynamics of the AJ vortex chain is parameterized  
by solutions of two coupled first order nonlinear differential equations which describe 
self-consistently the time dependence of the vortex velocity and the AJ core length.  
Exact formulas for the dc flux flow resistivity $R_f(H)$, and the nonlinear voltage-current 
characteristics are obtained. Dynamics of the AJ vortex chain driven by superimposed ac and 
dc currents is considered, and general expressions for a linear complex resistivity $R(\omega)$ and  
dissipation of the ac field are obtained. A flux flow resonance is shown to occur at 
large dc vortex velocities $v$ for which the imaginary part of $R(\omega)$ 
has peaks at the "washboard" ac frequency $\omega_0=2\pi v/a$, where $a$ is the inter vortex spacing.   
This resonance can cause peaks and portions with negative differential conductivity 
on the averaged dc voltage-current (V-I) characteristics. Ac currents of large amplitude cause 
generation of higher voltage harmonics and phase locking effects which manifest themselves in steps 
on the averaged dc I-V curves at the Josephson voltages, $n\hbar\omega/2e$ with $n=1.2, ...$. 
    
\end{abstract}

\pacs{PACES numbers: \bf 74.20.De, 74.20.Hi, 74.60.-w}]
\maketitle

\section{Introduction}

Mechanisms of current transport through grain boundaries (GB) in high-temperature 
superconductors (HTS) have attracted much attention because they reveal the d-wave 
symmetry of the HTS pairing\cite{sym} and determine the current-carrying capability of HTS 
materials\cite{dcl}. Unlike low-$T_c$ superconductors, GBs in HTS exhibit  
weak link behavior due to the exponential drop of  the local critical current density of a GB, 
$J_b=J_0\exp(-\vartheta/\vartheta_0)$, as the misorientation angle 
$\vartheta$ between the neighboring crystallites increases above $\vartheta_0\simeq 5-6^{\circ}$. 
The strong dependence 
of $J_b(\vartheta)$ on $\vartheta$ makes high-angle GBs crucial current-limiting defects  
in  HTS polycrystals\cite{dcl}. Since pioneering experiments of the IBM group\cite{gb}, much 
progress has been made in understanding the multiscale microstructure of GBs and its effect on 
their weak-link behavior \cite{gb1,gb2,gb3,rmp}, but primarily in the absence of a strong magnetic field H. 
Detailed atomic structure of GBs revealed by high-resolution electron microscopy 
have been used to determine local under doped states of GB, defect-induced suppression of 
superconducting properties at the nanoscale and controlled increase of $J_b$ by overdoping of 
GB\cite{dop1,dop2,dop3,dop4}. Recent models have also pointed out the importance of 
charging and strain effects which drive the HTS state at GB toward the metal insulator 
transition as $\vartheta$ increases\cite{mod,mh}.

At the same time, little is known about vortices on low-angle GBs, although it is the dynamics and pinning of 
the GB vortices, which mostly limit critical currents of HTS polycrystal in a magnetic field\cite{dcl}, 
and determine a microwave response of HTS films\cite{rf1,rf2,rf3,rf4,rf5}. Generally, pinning of vortices along GBs is weaker 
than in the grains, so GBs form a natural percolating network for preferential motion of vortices through a 
superconductor\cite{ag,aga,gc,diaz,ornl,anl,albr,hogg,claus}. Such percolating networks are not only 
characteristic of polycrystals, but represent a rather generic feature of vortex dynamics 
and pinning in superconductors. For instance, networks of easy flow vortex channels have been 
suggested by Kramer in early shear models of flux pinning\cite{kramer} and 
later discovered in molecular dynamics simulation of vortices in {\it random} pinning potential\cite{jensen,nori}, 
numerical simulations of time-dependent Ginzburg-Landau equations that describe moving vortex structure near twin 
boundaries\cite{crab} and observed by decoration \cite{dec} and Lorentz microscopy\cite{ton}. 
Matching effects in dynamics and pinning of mesoscopic vortex flow channels in artificial 
thin film superconducting structures have been extensively studied by Kes and co-workers\cite{kes1,kes2}.  
Many observable features of global current-voltage characteristics, magnetization, rf response 
and flux creep of HTS polycrystal may be due to dynamics and pinning of vortices in easy-flow channels, 
rather than stronger pinned vortices in the grains\cite{per1,per2,per3,per4,per5}. By contrast, GBs in low-$T_c$ materials 
do not block {\it macroscopic} currents, but can enhance flux pinning\cite{suenaga,ros} and play the role of  "hidden"  
weak links which strongly affect the vortex mass and viscosity 
\cite{ag} crucial for transport and microwave response of superconductors\cite{rf1,rf2}.

The behavior of vortices in easy flow channels on GBs in polycrystal is mostly determined by the structure 
of vortex cores which depends on the local depairing current density $J_b$ through a GB at 
the nanoscale of few current channels between the dislocations. 
The extreme sensitivity of $J_b(\vartheta)$ to the misorientation angle $\vartheta$ 
makes GBs in HTS a unique tool to trace a fundamental transition between the Abrikosov (A) and Josephson (J) vortices. 
As $\vartheta$ increases, $J_b(\vartheta)$ rapidly decreases, from the bulk depairing current density 
$J_d$ at $\vartheta\ll\vartheta_0$ down to much lower values $J_b\ll J_d$ at $\vartheta\gg\vartheta_0$. In turn,  
vortices on a GB evolve from the A vortices with normal cores at $\vartheta\ll\vartheta_0$ to intermediate Abrikosov vortices 
with Josephson cores (AJ vortices) \cite{ag} and then to the Josephson (J) vortices at higher $\vartheta$. 
There is no order parameter suppression in the AJ core, which is a phase kink whose length 
$l$ along GB is greater than the coherence length $\xi$, but shorter than the scale of circulating screening currents 
set by the London penetration depth $\lambda$. As $\vartheta$ further increases, the AJ vortices turn 
into J vortices in which both the Josephson currents and the magnetic field $H(x)$ vary on 
the same scale along GB set by the Josephson penetration depth $\lambda_J$\cite{barone,kkl}.
This continuous A to AJ vortex transition occurs as the spacing between GB dislocation cores becomes 
shorter than $\xi$, giving rise to a suppression of the amplitude $\Delta$ of the order parameter $\Psi=\Delta\exp(i\varphi)$ 
in current channels between dislocations \cite{mod}. Thus, a low-angle GB behaves as a high-$J_b$ 
superconducting-normal-superconducting (SNS) Josephson contact, for which the Josephson cores of the AJ vortices 
do not cause pair breaking effects responsible for the suppression of $\Delta$ in the normal A cores. Such contacts 
are described by integral equations of a nonlocal Josephson electrodynamics (NJE)
\cite{ag,gc,kkl,ivan,silin,alf,ms,ames,kuz} which account for the variations of phase difference 
$\theta(x)=\varphi_1-\varphi_2$ along a GB on any length scale greater than $\xi$. If $\theta(x)$ varies slowly 
on the scales $\sim\lambda$, the NJE equations reduce to the usual sine-Gordon equation for long Josephson 
junctions\cite{kkl,barone}. The key difference of the nonlocal approach from the local sine-Gordon theory  
is that the NJE equations can describe the AJ vortex core in the region of parameters where 
$J_b>J_d\xi/\lambda$, and $\theta(x)$ varies on the scale 
$l=\lambda_J^2/\lambda\simeq\xi J_d/J_b$ much shorter than the decay length $\lambda$ 
of the circulating supercurrents.

The importance of the Josephson nonlocality for thin films has been recognized long ago\cite{kkl,ivan}.
Because the penetration depth $\tilde{\lambda}=2\lambda^2/d$ increases as the film thickness $d$ 
decreases, the nonlocality condition $\lambda_J<\tilde{\lambda}$ can be fulfilled even for comparatively 
low-$J_b$ junctions. Independently, the NJE approach was developed for bulk superconductors to describe mixed AJ vortices 
on high-$J_b$ "hidden weak links", such as low-angle GBs in HTS and thin $\alpha$-Ti ribbons in NbTi \cite{ag}. A nonlocal 
generalization of the sine-Gordon equation was also considered in Ref. \onlinecite{silin}. It turns out that, in the 
strong nonlocality limit, the NJE equation reduces to the well-studied Peierls equation of dislocation theory, thus, the 
AJ single vortex solution \cite{ag} is similar to that for the core of an edge dislocation\cite{seeger}.   
The NJE equations have other exact solutions\cite{ag,silin,seeger,alfimov,let} for static and dynamic AJ vortex 
structures. Recently, the existence of AJ vortices in low-angle $YBa_2Cu_3O_7$ bicrystals was proven by transport 
measurements, using an exact expression for the flux flow resistivity of AJ vortices\cite{let}. The good agreement 
between the theory and experiment 
made it possible to extract the core length $l(T)$ and the intrinsic depairing current density $J_b(\vartheta)$ of a GB 
on a nanoscale of few dislocation spacings. The temperature dependence of $J_b\propto (T_c-T)^2$ extracted from these 
measurements does indicate the SNS coupling on GBs in HTS, in agreement with the model of Ref. \onlinecite{mod}.  

This paper presents a theory of a nonlinear flux flow of AJ vortices driven by 
dc and ac currents in a magnetic field. An exactly solvable model, that describes a chain of AJ 
vortices moving along a GB through the strongly pinned A vortex lattice in the grains, is proposed.    
This model of the overdamped vortex dynamics describes self-consistently  
both nonlinear dissipative processes in the AJ vortex cores and magnetic interaction between 
AJ vortices, showing how the distributions of circulating superconducting 
and quasiparticle currents change as a function of the vortex velocity.  
The AJ vortices exhibit many characteristic features of the dynamics of the periodic A vortex lattice, 
for example, viscous flux flow\cite{bs}, 
static and dynamic matching effects and Josephson-like voltage oscillations\cite{kes2,fiore,sh}. 
At the same time, the AJ vortices can also exhibit effects characteristic of 
the dynamics of short Josephson contacts in ac field, or J vortices in long Josephson junctions, 
for example, flux flow resonance\cite{est}, phase locking in ac field\cite{barone,ffo}, etc. 
Pronounced resonance effects occur if the strongly overdamped AJ structure is driven by 
superimposed ac and dc currents $J(t)=J_0+J_a\cos\omega t$ at $J_0>J_b$. It is shown that 
all this rich AJ vortex dynamics is described by two coupled first order nonlinear ordinary differential 
equations for the vortex velocity and the core size, for any time-dependent $J(t)$. These 
equations have the form of a complex resistively-shunted junction (RSJ) equation for a short Josephson contact. 

The paper is organized as follows. In Sec. II a qualitative description of 
length scales of vortices on a GB and their evolution with $\vartheta$ is given. 
In Sec. III the NJE equations and an exact solution 
that describes a chain of AJ vortices driven by an arbitrary ac current in a dc magnetic field are presented. 
In Sec. IV, the nonlinear $V-J$ characteristics and the flux flow resistivity 
of a GB in a magnetic field are calculated. 
In Sec. V a linear complex resistivity and rf dissipation are calculated for a chain of AJ vortices 
driven by superimposed ac and dc currents. A flux flow resonance is predicted. Sec. VI is devoted to nonlinear  
effects caused by superimposed ac and dc currents, in particular the averaged dc $V-J$ characteristics in the presence of 
an ac signal, generation of higher harmonics and phase locking 
effects.  Sec. VII concludes with a discussion of the obtained results.     

\section{Vortex length scales on grain boundaries in a magnetic field.}

The results of this paper are independent of the detailed atomic structure of GB\cite{gb,gb1,gb2}, 
so we consider a simplest planar [001] tilt GB 
between two crystallites misoriented by the angle $\vartheta$.  Such low-angle GB 
can be regarded as a periodic chain of edge dislocations spaced by $d_0=b/2\sin(\vartheta/2)$, 
where b is the Burgers vector\cite{gb2}. Because of the proximity of the HTS state to the 
antiferromagnetic metal-insulator transition, regions of size $\simeq b$ near dislocation cores 
are driven into insulating state by local nonstoichiometry, strains and charging effects\cite{mod}.  
As $\vartheta$ increases, the spacing  
$d_0(\vartheta)$ decreases, becoming smaller than the zero-T coherence 
length $\xi_0$ and the in-plane Debye screening length at the angle $\vartheta_0\simeq 4-6^{\circ}$. 
For $\vartheta>\vartheta_0$, the proximity effect, strain and charge coupling cause suppression of $\Delta$ 
between dislocation cores, which becomes more pronounced as $\vartheta$ increases. In this model  
a GB thus behaves as a SNS Josephson contact, whose critical 
current density $J_b(\vartheta)$ decreases nearly exponentially   
as $\vartheta$ increases with $J_b(\vartheta)\sim J_d$ at $\vartheta <\vartheta_0$. 

A magnetic field $H$ above the lower critical field $H_{c1}$ produces A vortices 
in the grain, and vortices of different character on the GB, depending on the ratio $\xi/d_0$. 
For $\theta\ll \theta_0$, the GB vortices are A vortices with normal cores 
pinned by GB dislocations \cite{diaz}.  As $\vartheta$ further 
increases, a GB exhibits a continuous transition from 
metallic to tunneling behavior above $\vartheta>\vartheta_0$, similar to high-$J_b$ SNS Josephson 
junction\cite{bkt} for which the normal core of A vortices disappears if $\Delta$ on the junction drops below a 
critical value\cite{kkl}. For vortices on a GB, the normal core disappears quite naturally because 
of the exponential decrease of $J_b(\vartheta)$. Since vortex currents must cross the GB which can only sustain 
$J_b$ much smaller than the depairing current density $J_d$, 
the modulus $\Delta$ of the order parameter is unaffected by vortex cores. As a result, 
the normal A core turns into a Josephson core in which the phase difference  
$\theta(x,t)$ on GB varies by $2\pi$ over the length $l$ along GB, but the 
amplitude $\Delta$ is independent of $x$.  The phase core length $l\simeq \xi J_d/J_b$ 
is greater then $\xi$, but smaller then the London penetration depth $\lambda$, if 
$J_b>J_d/\kappa$, where $\kappa=\lambda/\xi\simeq 10^2$ is the Ginzburg-Landau parameter \cite{ag}. 
As $\vartheta$ increases, the core length $l(\vartheta)\simeq\xi J_d/J_b(\vartheta)$ increases, so  
the GB vortices evolve from A vortices for $\vartheta\ll\vartheta_0$ to mixed AJ vortices at 
$J_d<J_b(\vartheta)<J_d/\kappa$. The AJ vortices turn into J vortices at higher angles, 
for which $l$ becomes greater than $\lambda$ if   
	\begin{equation}	%1
	\vartheta>\vartheta_J\simeq\vartheta_0\ln(\tilde{\lambda}/\xi ),
	\label{tj}
	\end{equation}
where $\tilde\lambda$ is the magnetic screening length. 
In bulk samples $\tilde\lambda$ is the London penetration depth $\lambda$, 
while a thin film of thickness $d<\lambda$ the magnetic screening length is 
$\tilde\lambda=2\lambda^2/d$.  For bulk samples, Eq. (\ref{tj}) yields $\theta_J\simeq 23^{\circ}$ if $\theta_0=5^{\circ}$, 
$\kappa=100$, and $J_0=J_d$. For a film with $d\ll\lambda$, the AJ region expands considerably, 
for example, if $d=0.1\lambda$, then $\vartheta_J\simeq 38^{\circ}$. 
Therefore, the AJ vortices exist in a rather wide range of misorientations, 
$\vartheta_0<\vartheta<\vartheta_J\simeq 22-40^{\circ}$ which comprises the crucial region 
of the exponential drop of $J_b(\vartheta)$. In this region the in-field current transport through GBs 
in HTS is determined by dynamics and pinning of AJ vortices. 

Due to the lack of the order parameter suppression in the AJ cores, AJ vortices can be 
described by the NJE theory which regards the GB as a Josephson 
contact whose $J_b(\vartheta)$ can be tuned in a very broad range by varying the 
misorientation angle  $\vartheta$. Once the AJ core size $l=\xi J_d/J_b$ exceeds 
the scales set by the coherence length, the Debye screening length and the dislocation spacing, the 
structure of AJ vortices is entirely determined by the electrodynamics of currents circulating 
around the cores,  regardless of the atomic structure of GB, pairing 
mechanisms and the symmetry of the order parameter. The 
current distribution in the AJ vortex is described by the universal Josephson 
and London equations, while the GB imposes a boundary condition of 
current continuity for the sum of the Josephson, quasiparticle and displacements current densities 
$J=J_b\sin\theta(x,t)+V/R + C{\dot V}$ crossing the GB, where  
$V$ is the voltage on a GB, $R$ and $C$ are the quasiparticle resistance and 
capacitance per unit area of a GB, respectively. 

The structure of a vortex on a GB is determined by the 
ratio $\kappa_b = {\tilde\lambda}/l$ reminiscent of the GL parameter $\kappa=\lambda/\xi$ for the A vortices. Here 
$\kappa_b=(\lambda/\lambda_J)^2\simeq \kappa J_b/J_d$ for a bulk sample, $\kappa_b=d\lambda^3/\lambda_J^2d$ for 
a film of thickness $d<\lambda$.  The case $\kappa_b\ll 1$ corresponds to the local 
relation $H(x)=\phi_0\partial_x\theta/4\pi\lambda$ characteristic of low-$J_b$ high-angle GB, for which 
both $\theta(x)$ and $H(x)$ vary on the same spatial scale.  Such GBs exhibit the J 
vortices of length $\lambda_J$ described by the sine-Gordon equation\cite{barone}. However, the low-angle GB 
(especially in thin films) correspond to $\kappa_b>1$ in which case the relation between $\theta(x)$ and $H(x)$ is nonlocal, 
and $\theta(x)$ and $H(x)$ vary on essentially different spatial scales $l=\xi J_d/J_b$ and $\lambda$, respectively. 
This gives rise to the mixed AJ vortices described by integral NJE equations\cite{ag,gc,kkl,ivan,silin,alf,ms,ames,kuz} 
which reduce to the sine-Gordon equation if $\theta(x)$ 
varies weakly on the scales $\sim\lambda$. The NJE equations provide a universal description of vortex structures 
on GB for which all microscopic details are hidden in the intrinsic parameters $J_b$ and $R$ of a GB. These 
parameters are very difficult to calculate, given the present state of the microscopic 
theory of HTS, but they can be extracted from resistive measurements on HTS bicrystrals with the help of exact 
NJE solutions that describe the flux flow resistance of moving AJ vortices \cite{let}.    
      
In a strong magnetic field $H\gg H_{c1}$, the A vortex spacing $a=(\phi_0/H)^{1/2}$ is shorter than 
$\lambda$, thus the relation between $\theta(x,t)$ and $H(x,y)$ is 
always nonlocal, regardless of the value of $J_b$.  The AJ vortex chain is then has two length scales: 
the core size $l>\xi$ and the 
inter vortex spacing $a(H)$. Both lengths $a$ and $l$ become comparable at a characteristic field 
$H_0\sim \phi_0/l^2\sim (J_b/J_d)^2H_{c2}$ much smaller then the upper critical field $H_{c2}$. 
Thus, unlike the A lattices, there is a wide field region $H_0<H<H_{c2}$ in which the AJ vortex cores 
overlap, but the bulk superconductivity persists. 

The larger core of AJ vortices leads to their weaker 
pinning along a GB, which thus becomes a channel for motion of AJ vortices between pinned 
A vortices in the grains \cite{ag} (Fig. 1). This gives rise to an extended linear region in the $V-I$ characteristic 
of a polycrystalline HTS that is dominated by motion of AJ vortices along GBs \cite{dop3,diaz,ornl,anl,albr,hogg,claus}. 
Pinning of AJ vortices results from interaction of the AJ phase core with structural inhomogeneities of GB 
and the magnetic interaction of AJ vortices with more strongly pinned A vortices in the grains. If the periods of 
the AJ vortices and bulk A vortices are slightly different, the AJ vortex chain breaks into 
commensurate domains (in which A and AJ periods coincide) separated by domain walls. 
This behavior, characteristic of commensurate-incommensurate transitions\cite{pt1,pt2}, was observed in 
molecular-dynamics simulations of A vortices in artificial flux flow channels \cite{kes2} for which the  
width of the domain walls (dislocations) considerably exceeds the inter vortex spacing. In this case   
the pinning of domain walls by the intrinsic Peierls potential is exponentially weak, so the depinning 
critical current $I_{gb}(H)$ is most likely due to macroscopic variations of superconducting 
properties along GB, for example, facet structures which cause significant peaks in $I_{gb}(H)$ if 
the A vortex spacing is commensurate with the facet period\cite{cai}, strains and local nonstoichiometry 
along GB\cite{dcl}, etc.     

At low field, only a single AJ vortex row 
moves along GB, while the A vortices in the grains remain pinned\cite{gc}. At higher field the 
moving AJ vortices start dragging neighboring rows of A vortices in a flux flow channel along GB. 
The field $H_1$ below which only a single AJ vortex row moves along the GB, 
can be estimated from the condition that the pinning force 
$f_m$ of AJ vortices due to their magnetic interaction with 
fixed A vortices equals the intergrain pinning force $\phi_0J_c/c$. 
The pinning force of AJ vortices is the maximum gradient of the 
magnetic energy $f(x)=-\phi_0\partial_x H(x)/4\pi$, where  
$H(x)=B+\Delta H\cos(2\pi x/a)$ is the local field produced by the fixed A vortex lattice along GB, 
$\Delta H = \phi_0e^{-2\pi u/a}/\pi\lambda^2$ is the amplitude of the oscillating 
part of the local field $H(x)$ due to the periodicity of the A lattice, and 
$u\sim a$ is the spacing of the first A vortex row from GB\cite{tern}. 
Therefore,
	\begin{equation}	%2
	H_1\simeq \Bigl[\frac{2\pi\lambda^2J_c}{c\sqrt{\phi_0}}\Bigr]^2\exp(\frac{4\pi u}{a}).
	\label{h1}
	\end{equation}
For $u=a$, $\lambda = 2000\AA$, and $J_c=10^5-10^6\mbox{A/cm}^2$, Eq. (\ref{h1}) yields 
$H_1\simeq 0.1-10$T. Note that the essential dependence of the transition field $H_1$ 
on the bulk $J_c$ indicates that the region $0<H<H_1$ can be significantly widen by irradiation 
which increases $J_c$ while weakly affecting GB properties \cite{let}. 
In addition, $H_1$ is very sensitive to the position $u$ of the first vortex row which can be strongly 
affected by the GB microstructure, facets, long-range $T_c$ variations due to strains, etc. 

In this paper we neglect the pinning of vortices on a GB, assuming that 
the driving current $J(t)$ is higher than the depinning current of the AJ vortices, but lower than $J_c$ of the A 
vortices in the grains. This behavior has been observed on HTS bicrystals in a wide 
region $H<H_1$ of magnetic fields\cite{dop3,diaz,ornl,anl,albr,hogg,claus,let}. Under the ac current, pinning effects  
in electromagnetic response weaken even more as the ac frequency $\omega$ exceeds   
a characteristic depinning frequency\cite{gitl,cofclem}. 
Furthermore, low-angle GBs can be regarded as overdamped Josephson contacts for which the 
displacements currents can be neglected. Indeed, the overdamped state corresponds to 
$\omega\ll \omega_c=(CR)^{-1}\sim 4\pi/\delta\rho\delta\epsilon$, where $\delta\rho$ and $\delta\epsilon$ are 
the excess resistivity and dielectric susceptibility on GB. Because for low-angle GB the dielectric dislocation 
cores do not overlap, $\delta\rho\sim\rho_n$, and $\delta\epsilon\sim 1$, where $\rho_n$ is the normal state 
resistivity at $T_c$. Thus, the condition $\omega\ll\omega_c$ always holds for $\omega$ smaller than 
the superconducting gap (see also Refs. \onlinecite{rf3,rf4}). We also neglect the time dispersion of the 
the GB resistance $R$ and contribution of bulk quasiparticles, adopting the simplest, frequency-independent 
$R$ in the framework of a standard RSJ model\cite{barone}.

\section{General dynamic equations.}

The NJE equations for current-driven vortex structures on an overdamped Josephson contact are \cite{ag,aga,gc}  
	\begin{eqnarray}	
	H=\frac{\phi_0}{(2\pi\lambda)^2}\int_{-\infty}^{\infty}
	\!\!\theta^{\prime}(u)K_0\Bigl[\frac{\sqrt{y^2+(x-u)^2}}{\lambda}\Bigr]du  
	+B_v, 
	\label{h} \\		%3
	\tau\dot\theta=\frac{l}{\pi}\int_{-\infty}^{\infty}
	\!\!\theta^{\prime\prime}(u)K_0\bigl(\frac{x-u}{\lambda}\bigr)du-
	\sin\theta + \beta \label{t},
	\label{th}\\		%4
	l=c\phi_0/16\pi^2\lambda^2J_b , 
	\qquad\qquad\tau = \phi_0/2\pi cRJ_b. 
	\label{const}		%5
	\end{eqnarray}
Here the overdot and the prime denote differentiation with respect to time and the coordinate $x$ along GB,  
$\beta(x,t)=J(x,t)/J_b$, ${\bf J}=(c/4\pi)\nabla\times {\bf B_v}$ is the current density 
across GB induced by bulk vortices, 
	\begin{equation}
	B_v(x,y)=\frac{\phi_0}{2\pi\lambda^2}\sum_nK_0\Bigl[\frac{|{\bf r}-{\bf r_n}|}{\lambda} \Bigr],
	\label{bv}		%6
	\end{equation} 
where ${\bf r_n}$ is the position of the n-th A vortex, $K_0(x)$ is a modified Bessel function,  
$\phi_0$ is the flux quantum, and $c$ is the speed of light. The first term in the right-hand side 
of Eq. (\ref{h}) describes the magnetic field produced by all currents circulating near GB, 
and the second term represents the contribution of bulk vortices without GB.  To provide the boundary 
condition for ${\bf J}(x,y)$ on a GB, the phase difference $\theta(x,t)$ must satisfy Eq. (\ref{t}), 
which results from the current continuity condition, $cH^{\prime}/4\pi=J_b\sin\theta + \hbar\dot\theta/2eR$.  
Eqs. (\ref{h})-(\ref{t}) describe spatial variation of $\theta(x,t)$ and $H(x,y,t)$ on any scale greater than 
$\xi$, irrespective of the microscopic mechanisms of current transport through the GB. The only assumption 
$J_b\ll J_d$ behind Eqs. (\ref{h})-(\ref{t}) ensures the lack of the order parameter suppression by currents 
flowing through GB. The geometry of the sample manifests itself 
in the long-range asymptotics of the kernel in Eq. (\ref{th}) at $|{\bf r}-{\bf r'}|>\lambda$. For instance, 
Eqs. (\ref{h})-(\ref{t})  correspond to an infinite GB in a parallel field. More complicated 
expressions for the kernel $\tilde{K}[(x-u)/\tilde{\lambda}]$ have been obtained for thin films in a 
perpendicular field\cite{ivan,ms,ames}, and slabs in parallel\cite{alf} and 
perpendicular\cite{kuz} fields.   
   
In the vicinity of the AJ cores, $r\ll\lambda$, and also in the high-field limit, 
$H\gg H_{c1}$, Eq. (\ref{th}) acquires a simple universal form independent of the 
sample geometry. This universality results from the fact that for $H\gg H_{c1}$, the derivative 
$\theta^{\prime\prime}(u)$ in Eq. (\ref{t}) rapidly oscillates over the inter vortex spacing 
$(\phi_0/H)^{1/2}\ll \lambda$. In this case the main contribution to the integral comes 
from the region $|x-u|<\lambda$, where the Bessel function $K_0(x)$ in Eq. (\ref{t}) can be 
replaced by its expansion $K_0(x)\simeq -\ln(x)$ at small $x$. Thus, the equation for 
$\theta$ becomes  
	\begin{equation}
	\tau\dot\theta=\frac{l}{\pi}\int_{-\infty}^{\infty}
	\frac{\theta^{\prime}(u)du}{u-x}-\sin\theta +\beta, 
	\label{nj} 		%7
	\end{equation}   
For other geometries, $K_0[(x-u)/\lambda]$ in Eq. (\ref{th}) should be replaced with the appropriate 
kernel $\tilde{K}[(x-u)/\tilde{\lambda}]$ which always has a logarithmic singularity at 
$x=u$, and a geometry-dependent nonsingular part, $\tilde{K}_{reg}(x,u)$: 
	\begin{equation}
	\tilde{K}(|x-u|/\tilde{\lambda})=-\ln |(x-u)/\tilde{\lambda}| + \tilde{K}_{reg}[(x-u)/\tilde{\lambda}].
	\label{tild}		%8
	\end{equation}
This general behavior of $\tilde{K}[(x-u)/\tilde{\lambda}]$ is illustrated in Appendix A where  
$\tilde{K}[(x-u)/\tilde{\lambda}]$ for a thin film ($d\ll\lambda$), is considered. Because of the rapid oscillations 
of $\theta^{\prime\prime}(u)$ at $H\gg H_{c1}$, the main contribution to the integral in Eq. (\ref{th}) comes from 
the narrow region around $u=x$, so neither $\tilde{K}_{reg}[(x-u)/\tilde{\lambda}]$, nor the screening length 
$\tilde{\lambda}$ contribute to Eq. (\ref{nj}). This feature of Eq. (\ref{th}) reflects the 
physical fact that the distribution of currents near the 
AJ cores is unaffected by the London screening. Indeed, the Green function 
$\tilde{K}[(x-u)/\tilde{\lambda}]$ is proportional to the single-vortex London solution $H(r)$, so 
the difference between $\tilde{K}[(x-u)/\tilde{\lambda}]$ for bulk samples and thin films is basically the same as 
between $H(r)$ for the A vortex and the Pearl vortex\cite{pearl}, respectively. Both vortices  
have the same distributions of currents $J(r)$ near the core, $r<\lambda$, 
but very different asymptotics of $J(r)$ for $r>\lambda$.  Thus, the universal Eq. (\ref{nj}) 
describes the distributions of the phase difference $\theta(x)$ in the AJ vortex cores and circulating 
supercurrents on the scales $<\tilde{\lambda}$ away from the cores where the London screening is inessential.

The driving parameter $\beta = \beta_0 + \delta\beta (x)$ is a sum of the constant transport current 
$\beta_0$ due to the gradient of the A vortex density in the grains, and an oscillating component 
$\delta\beta(x)$ due to the discreteness of the A vortex lattice. The term $\delta\beta(x)$ gives rise 
to a critical current $\beta_c$ through GB due to pinning of AJ vortices by A vortices in the grains \cite{gc}.  
In this paper we consider a rapidly moving AJ structure in the flux flow state, $\beta\gg\beta_c$, 
for which the pinning term $\delta\beta(x)\ll 1$ can be neglected, and  
$\beta(x)$ be replaced by $\beta_0(t)$.  
As shown in Appendix B, the nonlinear Eq. (\ref{nj}) then has the following {\it exact} 
solution that describes a stable periodic vortex structure:
	\begin{equation}
 	\theta=\pi+\gamma+2\tan^{-1}[M\tan k(x-x_0)/2], 
	\label{theta}		%9
	\end{equation}
where $\gamma(t)$, $M(t)$ and the vortex velocity $v(t)=\dot{x}_0$ 
depend only on t and obey the following equations
	\begin{eqnarray}
	\tau{\dot\alpha}+\sinh\alpha\cos\gamma=\sqrt{h}, 
	\label{al}\\		%10
	\tau{\dot\gamma}+\sin\gamma\cosh\alpha=\beta_0(t),
	\label{ga}\\		%11 
	k\tau v=-\sin\gamma\sinh\alpha.
	\label{u}		%12
	\end{eqnarray} 
Here $h = (kl)^2$ is the dimensionless magnetic field, 
the wave vector $k= 2\pi/a$ defines the period $a$ of the AJ structure, and 
	\begin{equation}
	\sinh\alpha=2M/(M^2-1).
	\label{sh}		%13
	\end{equation}
Using the complex variables $z=\gamma+i\alpha$ and $f=\beta_0+i\sqrt{h}$, 
Eqs. (\ref{al})-({\ref{u}}) can be written in a more compact form
	\begin{equation}
	\tau{\dot z}+\sin z=f,\qquad
	k\tau v=\mbox{Im}\cos z. 
	\label{comp}		%14
	\end{equation}
The equation for the complex  "phase" $z(t)$ has the same form as the  
usual RSJ equation for the phase difference on an overdamped point contact. 

Eqs. (\ref{theta})-(\ref{u}) were obtained in Appendix B by the Hilbert transform, 
which was used to obtain static periodic solutions of the Peierls equation (\ref{nj}) 
in the dislocation theory\cite{seeger}, and then employed to describe 
AJ structures \cite{alfimov}.  It is instructive to derive 
Eqs. (\ref{theta})-(\ref{u}) using a more transparent approach, starting from the 
basic London equation  
	\begin{equation}
	H-\lambda^2\nabla^2H=\phi_0\theta(x)\delta(y)/2\pi,
	\label{london}		%15
	\end{equation}
where $\theta(x)$ is determined by a particular vortex structure on a GB. 
Since screening does not affect the AJ cores and current 
distribution on the scales $<\lambda$ away from GB, Eq. (\ref{london}) reduces 
to the Laplace equation $\nabla^2H=0$ supplemented by the boundary condition for the tangential and 
normal components of the current density on GB:
	\begin{eqnarray}
	\partial_yH(x,+0)-\partial_yH(x,-0)=-\phi_0\theta'(x)/(2\pi\lambda^2),
	\label{bctan} \\		%16
	J_y(x,\pm 0)=-\frac{c}{4\pi}H'(x, \pm 0)=J_b\sin\theta+\frac{\phi_0\dot\theta}{2\pi cR}-J.
	\label{bcond}		%17
	\end{eqnarray}
If screening is inessential, $H(x,y)$ becomes a potential field, which in some cases can 
be found directly using the theory of analytic functions. For instance, $H(x,y)$ produced by the 
periodic AJ vortex structure in the {\it upper} half-plane $y>0$, is given by the following ansatz 
	\begin{equation}
	H=\frac{\phi_0}{2\pi\lambda^2}\mbox{Re}\ln\sin [x-x_0(t)+i(|y|+y_0(t))]\frac{k}{2},
	\label{he}		%18
	\end{equation} 
which describes the field produced by a chain of fictitious A vortices displaced by  
$y=-y_0$ away from GB in the {\it lower} half-plane (we omit a constant term in $H(x,y)$ inessential 
for $|y|<\lambda$). Eq. (\ref{he}) 
is an exact solution of the Laplace equation $\nabla^2H=0$. 
It turns out that $y_0(t)$ can be chosen  such that $H(x,y)$ also satisfies the boundary 
conditions (\ref{bctan}) and (\ref{bcond}) provided that $\theta(x,t)$ is given by Eq. (\ref{theta}) while  
$\gamma$, $\alpha$, and $x_0(t)$ obey Eqs. (\ref{al})-(\ref{u}). To show that we first observe that 
	\begin{eqnarray}
	\partial_xH=\frac{k\phi_0}{4\pi\lambda^2}\frac{\sin\zeta}
	{[\cosh k(y-y_0)-\cos\zeta]},
	\label{hpr} \\		%19
	\partial_yH=-\frac{k\phi_0}{4\pi\lambda^2}\frac{\mbox{sign}(y)\sinh\alpha}
	{[\cosh k(y-y_0)-\cos\zeta]},
	\label{hper}		%20
	\end{eqnarray}
where $\zeta = [x-x_0(t)]k$. In turn, Eqs. (\ref{theta}) and (\ref{sh}) yield 
	\begin{eqnarray}
	\theta'=\frac{k\sinh\alpha}{\cosh\alpha-\cos\zeta},
	\label{tprime} \\	%21 
	{\dot\theta}={\dot\gamma}-\frac{{\dot\alpha}\sin\zeta+k{\dot x_0}\sinh\alpha}{\cosh\alpha-\cos\zeta},
	\label{tdot} \\		%22
	\sin\theta=\frac{(1-\cosh\alpha\cos\zeta)\sin\gamma-\sinh\alpha\sin\zeta\cos\gamma}
	{(\cosh\alpha-\cos\zeta)}
	\label{sint}		%23 
	\end{eqnarray}
Now $y_0$ can be chosen such that the denominators of Eqs. (\ref{hpr})-(\ref{sint}) 
would coincide at $y=0$: 
	\begin{equation}
	ky_0=\alpha = \ln[(M+1)/(M-1)].
	\label{kya}		%24
	\end{equation}  
Eqs. (\ref{hper}) and (\ref{tprime}) with $ky_0=\alpha$ automatically satisfy the first boundary condition 
(\ref{bctan}).  Furthermore, substituting Eqs. (\ref{hpr})-(\ref{kya}) into the second 
boundary condition (\ref{bcond}) reduces the latter to the form: $C_1(t)\cos k\zeta+C_2(t)\sin k\zeta + C_3(t)=0$. 
The self-consistency conditions $C_i(t)=0$ are satisfied only if $\theta(t)$, $\alpha(t)$, and $x_0(t)$ do obey 
the dynamic equations (\ref{al})-(\ref{u}).	

Fig. 1 shows the current streamlines calculated from Eq. (\ref{he}), which has
a clear interpretation similar to that of a single AJ vortex\cite{ag}. 
Namely, the current streamlines described by Eq. (\ref{he}) in the upper 
half-plane $y>0$ coincide with those produced by a chain of moving fictitious A vortices 
displaced by $y=-\alpha(t)/k$ away from GB. Likewise, the current streamlines in the lower  
half-plane $y<0$ coincide with those produced by a chain of fictitious A vortices 
displaced by $y=\alpha(t)/k$ away from GB. The resulting nonsingular field distribution 
$H(x,y)$ is an exact solution for moving AJ vortices, where the transverse displacement 
$y_0$ determines the AJ core size along GB. The time-dependent Cartesian coordinates 
$x_0(t)$ and $y_0(t)=\alpha(t)/k$ of these fictitious A chains obey Eqs. (\ref{al})-(\ref{u}). 
The transition from AJ to A vortices occurs as $J_b$ increases, reaching 
$J_b\simeq J_d$, while $y_0$ decreases down to $y_0\simeq\xi$. Likewise, the transition from AJ to J 
vortices occurs as $J_b$ decreases below $J_d/\kappa$, in which case $y_0$ becomes greater than 
$\lambda$.

The set of coupled ordinary differential equations (\ref{al})-(\ref{u}) describe  
the evolution of the AJ phase core length, the phase shift $\gamma(t)$, and 
the vortex velocity ${\dot x}_0(t)$ for any time-dependent transport current $\beta_0(t)$. 
These equations along with Eqs. (\ref{theta}) and (\ref{he}) determine 
distributions of the phase difference $\theta(x,t)$ and screening currents for an interacting moving 
AJ vortex chain, including nonlinear dissipative dynamic states caused by ac and dc driving currents 
of large amplitude, $J(t)\sim J_b$. Eqs. (\ref{al})-(\ref{u}) considerably simplify in the low-field limit, 
$a\to\infty$ for which Eq. (\ref{theta}) reduces to a superposition of 
independent single AJ vortex solutions\cite{ag}
	\begin{equation}
	\theta(x,t)=\gamma(t)+\sum_n\Bigl[\pi+2\tan^{-1}\frac{x-na-x_0(t)}{L(t)}\Bigr].
	\label{lim}		%25
	\end{equation}
Here $k = 2\pi/a\to 0$, $\alpha\to 2/M\to 0$, so the AJ core length $L(t)=\alpha/k$ 
is independent of the magnetic field, while $\alpha=Lk\to 0$.  If  
$\alpha\to 0$, Eqs.  (\ref{al})-(\ref{u}) turn into the following equations 
that describe a single AJ vortex \cite{aga}
	\begin{eqnarray}
	\tau{\dot\gamma}+\sin\gamma=\beta_0,
	\label{ef1} \\		%26
	\tau{\dot L}+L\cos\gamma = l,
	\label{ef2} \\		%27
	\tau{\dot x}_0=-L\sin\gamma.
	\label{ef3}		%28
	\end{eqnarray}
Unlike Eqs.  (\ref{al})-(\ref{u})), the equation for 
$\gamma$ is {\it decoupled}, from Eqs. (\ref{ef2}) and (\ref{ef3}). 
Thus, in the limit $a\gg L$, the dynamics of the core length $L(t)$ and 
the vortex velocity ${\dot x}_0$ is integrable for any given $\gamma(t)$ which 
in turn is determined by Eq. (\ref{ef1}) for any time-dependent $\beta_0(t)$. 
This result  is no longer valid if the interaction between AJ vortices 
at a finite $H$ is taken into account, when both $\gamma(t)$ and $\alpha(t)$ 
are determined self-consistently by Eqs. (\ref{al})-(\ref{u}).

Contribution of each AJ vortex to the "staircase" solution (\ref{lim}) gives  
a $2\pi$ phase shift along GB, thus the AJ vortex carries exactly one flux quantum $\phi_0$\cite{gc}.   
Generally, the inter vortex spacing $a(H)$ on GB is different from the period $(\phi_0/H)^{1/2}$ 
of the bulk A lattice, because of the reduced $H_{c1}$ on a GB\cite{ag}. In the low-field region, 
$H\sim H_{c1}$, the magnetic induction $B$ is very different from $H$, which causes a significant 
mismatch in the periods of AJ and bulk A vortex lattices.  However, for strong fields  
$H\gg H_{c1}$ considered in this paper, the difference between B and H  is negligible, so 
the AJ spacing $a(H)$ nearly coincides with the bulk one, $(\phi_0/H)^{1/2}$ as both are fixed by 
the same flux quantization condition. 

The averaged voltage $V$ on a GB produced by the moving AJ vortex structure is given by: 
	\begin{equation}
	V=\frac{\phi_0}{2\pi ca}\int_0^a{\dot\theta}dx.
	\label{volt}		%29
	\end{equation}  
Substituting Eq. (\ref{tdot}) into Eq. (\ref{volt}), we see that the term proportional to ${\dot\alpha}$ 
in $\dot\theta$ vanishes after integration, while the part of 
${\dot\theta}(x-x_0)$ proportional to ${\dot x}_0$ reduces to the full derivative $-{\dot x}_0\theta'$. 
Integration of $\theta'$ from $0$ to $a$ gives $2\pi$ due to the 
flux quantization condition\cite{gc}, $\theta(x+a)-\theta(x)=2\pi$, thus
	\begin{equation}
	V=\frac{\phi_0}{2\pi c}(\dot\gamma -kv).
	\label{vol}		%30
	\end{equation}
Here the first term in the parenthesis describes the quasiparticle component of 
$V$, and the second term results from vortex motion. Eq. (\ref{vol}), along with  
Eqs. (\ref{al})-(\ref{u}), determine the nonlinear electromagnetic response 
of the moving AJ vortex chain.  These equations can be presented in different forms, 
depending on the way the external drive is applied.  For example, the solutions of 
Eqs. (\ref{al})-(\ref{u}) determine the voltage (\ref{vol})  for any time-dependent current $J(t)$. 

Another regime corresponds to the rf response, when it is 
the voltage $V(t)$ (rather than the current density $J(t)$) on a GB which is fixed by 
an external rf source.  In this case it is convenient to express the vortex velocity via  
$V$ from Eq. (\ref{vol}) and subtract Eq. (\ref{ga}) from Eq. (\ref{uu}). Then  
the equations for $\alpha(t)$, $\gamma(t)$ and $\beta(t)$ in the fixed voltage mode take the form
	\begin{eqnarray}
	\tau{\dot\alpha}+\sinh\alpha\cos\gamma=\sqrt{h}, 
	\label{aal}\\		%31
	\tau{\dot\gamma}+\sin\gamma\sinh\alpha=u,
	\label{uu} \\		%32
	\beta=u+e^{-\alpha}\sin\gamma,
	\label{subtr}		%33
	\end{eqnarray}
where $u=V/RJ_b$ is the dimensionless voltage on a GB.

\section{Nonlinear steady-state flux flow}

The steady-state propagation of the AJ vortex chain driven by the 
constant current $\beta_0$ is described by  Eqs. (\ref{theta}) and (\ref{al})-(\ref{u})  
with ${\dot \alpha}={\dot\gamma}=0$, whence $\tanh\alpha=-s\sqrt{h}/\beta_0$, $\tan\gamma=-s$,   
and $s\sqrt{h}=-\sin\gamma\sinh\alpha$. These equations give the dimensionless propagation velocity 
$s(\beta_0)=v/v_0$ in the form  
	\begin{equation}
	s^2=[\sqrt{(1-\beta_0^2+h)^2+4\beta_0^2h}-1-h+\beta_0^2] /2h.
	\label{sv}		%34	
	\end{equation}
Here $v_0=l/\tau$, $h=(2\pi l/a)^2$ is a dimensionless magnetic field. 
The limit $h\to 0$ corresponds to a single AJ vortex, for which 
both Eq. (\ref{sv}) and the steady-state Eqs. (\ref{ef1})-(\ref{ef3}) ($L\cos\gamma=l$, $\sin\gamma=\beta_0$ 
and $\tau v=-L\cos\gamma$), give $v=v_0\beta_0/\sqrt{1-\beta_0^2}$. The so-obtained single vortex velocity 
$v(J)$ diverges at $J\to J_b$, because the AJ core size $L=l/\sqrt{1-\beta_0^2}$ expands as $\beta_0$ increases\cite{ag}. 
The core expansion as the velocity $v(J)$ increases is characteristic 
of both AJ\cite{ag} and J\cite{kkl} vortices in the overdamped limit, unlike the Lorentz contraction of  
J vortices in the underdamped limit. However, as the AJ cores expand, they start overlapping, so  
the interaction between vortices at $J> J_b$ cannot be neglected even at low fields, $h\ll 1$. 
The importance of the interaction is apparent form the exact Eq. (\ref{sv}), which shows that the 
velocity $s(\beta_0)$ smoothly increases as $\beta_0$ increases and has no singularity 
for any nonzero $h$.     

The dc $V-J$ characteristic due to viscous motion of AJ vortices follows from Eq. (\ref{vol}) 
in which $\dot\gamma=0$, and $v(J)$ is given by Eq. (\ref{sv}). Hence  
	\begin{equation}
	V=\frac{V_0}{\sqrt{2}}\Bigl[\sqrt{(1-\beta_0^2+h)^2+4\beta_0^2h}-1-h+\beta_0^2\Bigr]^{1/2},
	\label{vj}		%35
	\end{equation}
where $V_0=RJ_b$. Eq. (\ref{vj}) can also be written in a more transparent form
	\begin{equation}
	J=\frac{V}{R}\Bigl[1+\frac{1}{H/H_0+(V/V_0)^2}\Bigr]^{1/2}.
	\label{jv}		%36
	\end{equation}  
The $V-J$ curve shown in Fig. 2 has two linear portions: a flux flow part, $V=R_fJ$ at $J\ll J_b$,  
and the quasiparticle Ohmic part, $V=RJ$ for $J\gg J_b$. In the crossover region, 
$J\sim J_b$, the $V-J$ curve becomes nonlinear, because of the AJ core  
expansion as $J$ increases.  Since the AJ vortex core overlap at 
$J>J_b$, the GB resistance approaches its maximum value $R$ at $J\gg J_b$.   

For $J\ll J_b$, Eq. (\ref{jv}) yields $V=R_fJ$, where $R_f=R\sqrt{h/(1+h)}$ is 
the flux flow resistivity of AJ vortices. If $H\gg H_{c1}$, then $h= (2\pi l/a)^2=H/H_0$, and 
	\begin{equation}
	R_f=\frac{R\sqrt{H}}{\sqrt{H+H_0}},\quad\quad H_0=\frac{\phi_0}{(2\pi l)^2}.
	\label{r}		%37
	\end{equation}    
At $H\ll H_0$, Eq. (\ref{r}) describes a chain of AJ vortices whose cores do not 
overlap. In this case $R_f(H)$ is similar to the 1D Bardeen-Stephen formula\cite{bs}, 
$R_{BS}\simeq R\sqrt{H/H_{c2}}$, except that in Eq. (\ref{r}) the core structure is taken into 
account exactly (see also Ref. \onlinecite{ag}). For $H>H_0\simeq (J_b/J_d)^2H_{c2}\ll H_{c2}$, 
the AJ cores overlap, and Eq. (\ref{r}) describes a crossover to a field-independent  
resistance $R$. This regime has no analogs for A vortices, whose normal overlap only at $H_{c2}$. 

It is interesting to compare Eq. (\ref{r}) to $R_f(H)$ for J vortices 
at $J\ll J_b$. It is known\cite{ls} that for small density of J vortices, $R_f(B)=RB/H_{c1J}$ 
is proportional to $B$, similar to the Bardeen-Stephen resistivity in which $H_{c2}$ is 
replaced by the Josephson lower critical field $H_{c1J}=\phi_0/\pi^2\lambda\lambda_J$. 
A general expression for $R_f(H)$ can be written in the following form (see Appendix C):
	\begin{equation}
	R_f(H)=RB(H)/H.
	\label{rfj}		%38
	\end{equation}
Here $B(H)$ is determined by the 
equilibrium magnetization the J vortex structure\cite{os}:
	\begin{equation}
	B=\pi^2H_{c1J}/4pK(p),\qquad H=H_{c1J}E(p)/p,
	\label{bhj}		%39
	\end{equation}
where $K(p)$ and $E(p)$ are the complete elliptic integrals of the first and the second kinds, 
respectively\cite{abr}, and $0<p<1$ is a continuous parameter. As follows from Eqs. (\ref{rfj}) and (\ref{bhj}), 
$R_f$ first increases linearly with B at $B\ll H_{c1J}$ and then approaches the constant 
value $R$ at $H\gg H_{c1J}$ as the J vortices overlap, and $B\to H$. By contrast, the saturation 
of $R_f(H)$ for the AJ vortices occurs at much higher fields $\sim H_0\gg H_{c1}$
for which the AJ vortex spacing $a=(\phi_0/B)^{1/2}$ becomes of order the AJ core size $l$. 
The dependence of $R_f(H)$ for J vortices looks rather different from the simple Eq. (\ref{r}) mostly because of the complicated 
relation  (\ref{bhj}) between B and H at low fields $H\simeq H_{c1J}$. In fact, Eq. (\ref{r}) also becomes more complicated 
at low H, of the order of the AJ lower critical field $H_{c1b}=(\phi_0/4\pi\lambda^2)[\ln(\lambda/l)+0.423]$ \cite{ag}, 
for which $H$ in Eq. (\ref{r}) should be replaced by the corresponding $B(H)$ dependence for AJ vortices \cite{alf}. 
For $H\gg H_{c1b}$, the induction B almost coincides with H, thus $R_f(H)$ for AJ vortices acquires  
the universal form (\ref{r}) independent of demagnetizing effects crucial at $H\sim H_{c1}$. 
The simplicity of Eq. (\ref{r}) is very convenient to extract intrinsic properties of GBs at the 
nanoscale from transport measurements of $R_f(H)$ in HTS bicrystals \cite{let}. 

\section{Linear ac response}

To obtain the dynamic linear resistance $R_\omega$ for a weak ac current, 
$J_a\cos(\omega t)\ll J_b$ superimposed on the dc current $J$, we calculate 
the amplitude of the induced ac voltage, $V_\omega$ from Eq. (\ref{vol}) in which  
$\gamma(t)$ and $v(t)$ are determined by Eqs. (\ref{al})-(\ref{u}) with
$\beta=\beta_0+\beta_a\exp(i\omega t)$. Setting  
$\alpha = \alpha_0+\delta\alpha$, $\gamma = \gamma_0+\delta\gamma$ and 
calculating the perturbations $\delta\alpha\ll 1$ 
and $\delta\gamma\ll 1$ induced by the ac current from the linearized Eqs. (\ref{al})-({\ref{u}}) yields 
the following general expression for the ac complex resistivity 
$R_\omega = V_\omega/J_a$ (see Appendix D):
	\begin{equation}
	\frac{R_\omega}{R}=\frac{\eta(h+u^2)/\sqrt{h}-\Omega^2+i\Omega(\eta+\sqrt{h})}
	{\Omega_0^2-\Omega^2+2i\eta\Omega}.
	\label{rdin}		%40
	\end{equation}
Here $u=V/J_bR$ is the dimensionless dc voltage on a GB,
$\Omega = \omega\tau$ is the normalized ac frequency, $\Omega_0$ and $\eta = \beta_0/s$ 
are the dimensionless flux flow resonance frequency and viscosity, respectively 
	\begin{equation}
	\eta = \sqrt{h+h/(h+u^2)},\qquad\quad \Omega_0=\sqrt{u^2+\eta^2},
	\label{eo}		%41
	\end{equation}

In the fixed current mode, it is convenient to express $\Omega_0$ and $\eta$ in terms of $\beta_0$:
	\begin{eqnarray}
	\Omega_0=[(1+h-\beta_0^2)^2+4h\beta_0^2]^{1/4},
	\label{omj}\\		%42
	\eta=[\sqrt{(1+h-\beta_0^2)^2+4h\beta_0^2}+1+h-\beta_0^2]^{1/2}/\sqrt{2}
	\label{etaj}		%43
	\end{eqnarray}
The dependencies of $\Omega_0$ and $\eta$ on $\beta_0$ are shown in Fig. 3. For $J<J_b$, 
$\Omega_0$ and $\eta$ practically coincide, but for higher $J$, the frequency $\Omega_0$ becomes 
much higher than the damping constant $\eta$.
In the following subsections various dynamic regimes described by Eq. (\ref{rdin}) are considered.

\subsection{Small dc vortex velocities}

For $v=0$, Eq. (\ref{rdin}) reduces to 
	\begin{equation}
	R_\omega = R +\frac{R_f-R}{i\omega\tau_f+1}.
	\label{rom}		%44
	\end{equation}
Here $R_f$ is the dc flux flow resistivity (\ref{r}), and $\tau_f$ 
is the flux flow relaxation time constant in the AJ vortex chain:
	\begin{equation}
	\tau_f=\frac{\tau}{\sqrt{1+H/H_0}}
	\label{tf}		%45
	\end{equation}
The Drude-like frequency dependence of $R_\omega$, results in the following retarded 
relation between the induced voltage $V(t)$ and the driving current $J(t)$  
	\begin{equation}
	V(t)=RJ(t)+(R_f-R)\int_{-\infty}^t e^{\frac{t'-t}{\tau_f}}J(t')\frac{dt'}{\tau_f}.
	\label{vt}		%46
	\end{equation} 
For example, after a jump-wise increase of $J(t)$ 
from $0$ to $J_0$, the steady-state flux flow sets in according to 
	\begin{equation}
	V(t)=J_0[R_f+(R-R_f)e^{-t/\tau_f}],\quad t>0,
	\label{vjz}		%47
	\end{equation}  
and $V=0$ for $t<0$. The discontinuity in $V(t)$ at $t=0$ disappears if a time dispersion 
of $R$, or a finite capacitance $C$ of the GB are taken into account. In the latter 
case $V(t)$ sharply increases from $0$ to $V(t)$ given by Eq. (\ref{vjz}) during a short time 
$\tau_i = RC$.  

The ac power $Q=(1/2)J_a^2Re(R_\omega)$, dissipated on a GB due to viscous flow of AJ vortices 
can be obtained from Eqs. (\ref{r}), (\ref{rom}), and (\ref{tf}) in the form 
	\begin{equation}
 	Q=\frac{RJ_a^2[\sqrt{h(1+h)}+(\omega\tau)^2]}{2[1+h+(\omega\tau)^2]}.
	\label{q}		%48
	\end{equation} 
For a fixed frequency, $Q(H)$ monotonically increases with $H$, approaching the quasiparticle limit 
$RJ_a^2/2$ for $H\gg H_0$, as shown in Fig. 4. In the steady state, $\omega\tau\to 0$, 
the power $Q(H)\propto \sqrt{h}/\sqrt{1+h}$ is 
simply proportional to $R_f(H)$. For finite frequencies, $Q(H)$ becomes finite even with no vortices $(H=0)$ 
due to quasiparticle ac ohmic currents through GB.

\subsection{Flux flow resonance for moving AJ vortices.}

Interaction of the moving AJ chain with the ac field can cause a resonance, 
if $\omega$ is close to the real part $\omega_0$ of the complex 
eigenfrequency $\omega_f$ which corresponds to the pole in $R_\omega(\Omega)$: 
	\begin{equation}
	\omega_f = \omega_0 + i\eta_0,
	\label{oi}		%49 
	\end{equation}
where the flux flow resonance frequency $\omega_0$ and the damping constant 
$\eta_0$ are given by
	\begin{eqnarray}
	\omega_0&=&\sqrt{(kv)^2+\eta_0^2},
	\label{od} \\		%50
	\eta_0&=&\frac{1}{\tau}\Bigl[\frac{H}{H_0}+\frac{1}{1+(v/v_0)^2}\Bigr]^{1/2}.
	\label{ed}		%51
	\end{eqnarray}   

For small dc velocity $v$, the AJ oscillations are strongly overdamped, $\omega_0\simeq \eta_0$, 
so no resonance peaks in $R_\omega(\Omega)$ occur for $J<J_b$, as evident from Fig. 3. 
However because the frequency $\omega_0$ increases, while the damping constant $\eta_0$ decrease as 
$v(J)$ increases, the flux flow resonance emerges at high dc driving currents $J >J_b$, for which $\omega_0\gg\eta_0$.  
For a given frequency $\omega$, the resonance occurs at  the vortex velocity $v_f$, for which 
	\begin{equation}
	\omega^2=(kv_f)^2 + \eta_0^2(v_f).
	\label{res}		%52
	\end{equation}
If $kv\gg\eta_0$, the resonance frequency $\omega_0$ approaches the "washboard " frequency\cite{fiore} 
for which the vortex velocity $v_f$ equals the phase velocity 
of the electromagnetic wave $\omega/k$ with the wave vector $k=2\pi/a$ of the AJ structure. 
Because $\Omega_0(\beta_0)$ has a minimum at $J\simeq J_b$ (see Fig. 3), the 
resonance condition (\ref{res}) at a given frequency $\omega$ can be satisfied either for one 
or two velocities $v_f$.  From Eqs. (\ref{od}), (\ref{ed}), and (\ref{res}), it follows that 
	\begin{equation}
	v_f=\Bigl[\frac{\Omega^2\pm\sqrt{\Omega^4-4h}}{2h}-1\Bigr]^{1/2}v_0.
	\label{vf}		%53
	\end{equation}  
The solution for $v_f$ exists above the threshold $\Omega >(4h)^{1/4}$, where the resonance can occur 
either at one or two different velocities $v_f$, depending on $\Omega$: 
	\begin{eqnarray}
	(4h)^{1/4}&<&\Omega<(1+h)^{1/2}\qquad \mbox{two}\;\; v_f 
	\label{ineq1} \\		%54
	\Omega &>& (1+h)^{1/2}\qquad \mbox{one}\;\; v_f .
	\label{ineq2}		%55
	\end{eqnarray}
At $\Omega=(1+h)^{1/2}$, the smaller resonance velocities vanishes as 
$v_f\propto (\sqrt{1+h}-\Omega)^{1/2}$. For $\Omega\gg h^{1/4}$, the 
larger resonance velocity approaches the material-independent "washboard" 
value $v_f=a\omega/2\pi$.

Near the resonance, $\omega\approx\omega_0$, at high vortex velocities ($s^2\gg 1, 
\: \Omega_0\approx s\sqrt{h},\:  \eta\approx \sqrt{h}\ll \Omega_0$),  Eq. (\ref{rdin}) yields  
	\begin{equation}
	\frac{R_\omega}{R}=1+\frac{1}{4\Omega_0(i\eta-\Omega + \Omega_0)}
	\label{resam}		%56
	\end{equation}
Eq. (\ref{resam}) describes a resonance line with a Lorentz peak in 
$\mbox{Im}R_\omega=-R/4s[h+(\Omega-\Omega_0)^2]$, as shown in Fig. 5. 
For $s^2\gg 1$, the amplitude of the peak decreases as $v(J)$ 
and H increase. The resonance is most pronounced if  
$J>J_b$ and $H\ll H_0$, while at smaller $J$ or higher H, the peak in $Im R_\omega(\Omega)$ 
disappears as the linewidth $\eta$ becomes of the 
order of the eigenfrequency $\Omega_0$.   

\section{Nonlinear response}

\subsection{Flux flow resonances on the dc V-J curve}

The flux flow resonance also manifests itself in the averaged dc voltage $V$ as a function of the dc current density. 
To calculate $V(J)$, it is convenient to use the complex representation (\ref{comp}) for $z=\gamma+i\alpha$, taking  
$z=z_0+\delta z_0+\delta z$, where $z_0$ is the dc solution without the ac field for which $\sin z_0=\beta+i\sqrt{h}$, 
and the oscillating correction $\delta z$ obeys the linearized equation 
	\begin{equation}
	\tau\delta\dot{z}+\cos z_0\delta z=\delta\beta. 
	\label{comd}		%57
	\end{equation}
Here $\delta\beta(t)=\beta_a\cos\omega t$, and $\delta z_0$ is a dc correction to $z_0$ due 
to the ac field, which is determined by Eq. (\ref{comp}) expanded to quadratic terms in $\delta z$:
	\begin{equation}
	2\delta z_0\cos z_0 = \langle\delta z^2\rangle\sin z_0,
	\label{zo}		%58
	\end{equation} 
where the angular brackets mean time averaging over the ac period $2\pi/\omega$. 
The averaged vortex velocity  $\langle v\rangle$ is given by the second of 
Eqs. (\ref{comp}) expanded to quadratic terms in $\delta z$ and linear terms in $\delta z_0$:
	\begin{equation}
	k\tau\langle v\rangle = Im\left(\cos z_0-\frac{\langle\delta z^2\rangle}{2\cos z_0}\right).
	\label{vs}		%59
	\end{equation}
The second term in the parenthesis describes the correction due to the ac field. The value 
$\langle\delta z^2\rangle$ can be calculated using the solution of Eq. (\ref{comd}),
	\begin{equation}
	\delta z=\frac{\beta_a}{\Omega^2+\cos^2z_0}(\cos z_0\cos\omega t+\Omega\sin\omega t),
	\label{soldz}		%60
	\end{equation}
which yields $\langle\delta z^2\rangle=\beta_a^2/2[\Omega^2+\cos^2z_0]$. Inserting this expression 
into Eq. (\ref{vs}) gives the averaged vortex velocity in the form  
	\begin{equation}
	\sqrt{h}\langle s\rangle = -(1+\Gamma_\omega)\sin\gamma_0\sinh\alpha_0,
	\label{sa}		%61
	\end{equation}
where the parameter $\Gamma_\omega$ quantifies the ac contribution 
	\begin{equation}
	\Gamma_\omega=\frac{J_a^2(4\eta^2+\Omega^2-\Omega_0^2)}
	{(2J_b\Omega_0)^2[(\Omega^2-\Omega_0^2)^2+4\eta^2\Omega^2]}.
	\label{gam}		%62
	\end{equation}
Eq. (\ref{sa}) for $\langle s \rangle$ is identical to the dc relation, $\tau kv=-\sin\gamma\sinh\alpha$, 
if $v$ is replaced by the effective velocity $\tilde{v} = v/[1+\Gamma_\omega(v)]$. Therefore,  
the averaged $J-V$ characteristics can be presented in the dc form (\ref{jv}) 
if $V$ is replaced by the effective voltage $\tilde{V}(V)$ as follows  
	\begin{eqnarray}
	J=\frac{\tilde{V}}{R}\Bigl[1+\frac{1}{H/H_0+(\tilde{V}/V_0)^2}\Bigr]^{1/2},
	\label{renv} \\		%63 
	\tilde{V}=V/[1+\Gamma_\omega(V)],
	\label{subst}		%64
	\end{eqnarray}
where $\Gamma(V)$ is determined by Eqs. (\ref{eo}) and (\ref{gam}). The $J-V$ characteristics 
described by Eqs. (\ref{renv}) and (\ref{subst}) 
$J_a$ are shown in Fig. 6 for different ac amplitudes $J_a$. For stronger ac signal, 
the $J-V$ curves can exhibit two maxima at the resonance voltages $V_f=\phi_0v_f/ca$, 
and portions with negative differential conductivity. As follows from Eqs. (\ref{ineq1}) 
and (\ref{ineq2}), there are either one or two resonance voltages, depending on the relation between 
$H$ and $\omega$. The increase of the magnetic field broadens the peaks in 
$J(V)$ which eventually disappear at higher $H$, because of the increase of the effective damping 
constant $\eta_0$ in Eqs. (\ref{oi})-(\ref{ed}).

\subsection{Ac dissipation}

The mean ac power $Q=\langle JV\rangle$ dissipated per unit area of a GB due to the ac voltage $V(t)=V_a\cos\omega t$ 
of large amplitude can be calculated by solving Eqs. (\ref{aal})-(\ref{subtr}) numerically. The situation 
simplifies for a low-frequency ac signal ($\omega\tau\ll 1$) for which $Q$ can be obtained using 
the quasistatic Eq. (\ref{jv}). For low fields $h\ll 1$ and moderate ac amplitudes $u_a=V_a/V_0<1$, 
the unity under the square root in Eq. (\ref{jv}) can be neglected, giving 
	\begin{equation}
	Q=\frac{\omega V_a^2}{2\pi R}\int_0^{2\pi/\omega}\frac{\cos^2\omega t dt}{\sqrt{h+u_a^2\cos^2\omega t}}.
	\label{qstat}		%65
	\end{equation}
This integral can be expressed in terms of the complete elliptic integrals $K(m)$ and $E(m)$\cite{abr}:
	\begin{eqnarray}
	Q=\frac{2J_b^2R_f}{\pi\sqrt{1+g^2}}[(1+g^2)E(m)-K(m)],
	\label{qstt} \\		%66
	m=g^2/(1+g^2),\qquad g=V_a/R_fJ_b.
	\label{mgstaff}		%67
	\end{eqnarray}
The power $Q(g)$ is a function of only one dimensionless parameter $g$ which 
includes both the ac amplitude and the magnetic field, as shown 
in Fig. 7. For a weak ac signal, $V_a\ll J_bR\sqrt{h}$, Eq. (\ref{qstt}) 
yields the quadratic dependence of $Q=V_a^2/2R_f$ on $V_a$, where $R_f=R\sqrt{h}$ 
is the dc flux flow resistivity at $h\ll 1$.  However, for stronger ac signals, $J_bR\sqrt{h}\ll V_a\ll J_bR$,  
Eqs. (\ref{qstat}) and (\ref{qstt}) yield the {\it linear} dependence  
	\begin{equation}
	Q=2V_aJ_b/\pi, 
	\label{linq}		%68
	\end{equation}
which is independent of the GB resistance R. This behavior is due to the AJ core expansion at large vortex 
velocities. For very high ac amplitude $V_a\gg RJ_b$, the full Eq. (\ref{jv}) should be used instead of 
Eq. (\ref{qstat}). In this case Eq. (\ref{jv}) yields $V=J/R$ during most part of the 
ac cycle, thus the ac power dissipated on the GB becomes 
quadratic in $V_a$, approaching the normal state limit, $Q\to V_a^2/2R$ (not shown in Fig. 7). 

\subsection{Generation of higher harmonics}

The nonlinearity of the electromagnetic response of the AJ structure driven 
by harmonic ac current $J(t)=J_a\cos\omega t$ gives rise to higher voltage harmonics
	\begin{equation}
	V(t)=\sum_{n=0}^{\infty}V_{2n+1}(\beta,\omega,h)\cos(2n+1)\omega t,
	\label{harmon}		%69
	\end{equation}
where the Fourier coefficients $V_{2n+1}$ can in principle be calculated from 
Eqs. (\ref{al})-(\ref{u}). The higher harmonics in $V(t)$ are most pronounced if the amplitude 
of ac AJ vortex displacements is maximum. For the overdamped dynamics 
considered in this paper, the amplitude of the AJ vortex oscillations decreases with  
$\omega$, so the higher harmonics in $V(t)$ are most pronounced for the 
quasi-static ac signal, $\omega\tau\ll 1$. In that case the coefficients, 
	\begin{equation}
	V_{2n+1}=\frac{2\omega}{\pi}\int_0^{\pi/\omega}dt V(\beta_a\cos\omega t)\cos(2n+1)\omega t 
	\label{vnstat}		%70
	\end{equation}
are independent of $\omega$, and $V[\beta(t)]$ is given by Eq. (\ref{vj}). 
For instance, the amplitude of the third harmonics $V_3$ for a small 
ac current $\beta_a=J_a/J_b\ll 1$ is obtained by expanding Eq. (\ref{vj}) up to cubic terms in $\beta$. 
This yields $V(t)\approx R_fJ_a\cos\omega t+V_3\cos 3\omega t$, where  
	\begin{equation}
	V_3=\frac{RJ_a^3}{8J_b^2}\frac{\sqrt{h}}{(1+h)^{5/2}}
	\label{v3}		%71
	\end{equation}
The amplitude of the third harmonics $V_3(h)$ has a maximum at $h=1/4$, as shown in Fig. 8. 
This field dependence reflects the increase of $V_3\propto \sqrt{h}$ at small $h$ proportional to the 
AJ vortex density, followed by the decrease of $V_3(h)$ at higher fields, for which the AJ cores start overlapping, 
and the $V(J)$ curve becomes Ohmic. This trend is characteristic of other higher order harmonics as well, 
whose amplitudes $V_{2n+1}(h)\propto\beta_a^{2n+1}$ strongly decrease with $n$ and $h$ for $h>1$.  

The maximum in $V_3(h)$ could be used for extracting the field $H_0$ 
and thus the AJ core size $l$ on a GB from the ac measurements. 
In that case the ac measurements of $V_3(H)$ may bring some advantages over the dc measurements of 
$H_0$ from the flux flow resistivity $R_f(H)$\cite{let}, since the maximum in $V_3(H)$ occurs at 
the field $H_0/4$ independent of the quasi-particle resistivity $R$, while the extracting of 
$l$ from $R_f(H)$ requires a two-parameter fit for $H_0$ and $R$.

\subsection{Phase locking and quasi-steps on $J-V$ curves}

AJ vortices in superimposed ac and dc fields can exhibit phase locking effects at ac large amplitudes $\beta_a$ if 
the dc voltage $V$ on a GB is commensurate to the Josephson voltage $V_\omega=\hbar\omega/2e$, 
where $n$ is any integer, and $e$ is the electron charge.
The electromagnetic response of a GB biased by a dc voltage $V$ superimposed on ac  
voltage $V_a\cos\omega t$, is described by Eqs. (\ref{aal})-(\ref{subtr}) 
with $u(t)=u+u_a\cos\omega t$, where $u=V/RJ_b$, and $u_a=V_a/RJ_b$. To calculate the dc $J-V$ characteristic 
averaged over the ac oscillations with the account of the phase locking of $\gamma(t)$ onto the ac field, 
we use an approach similar to that for Shapiro current steps on $J-V$ curves of small Josephson 
junctions \cite{ffo,barone}. In this case   
	\begin{equation}
	\gamma=\psi+n\omega t+q\sin\omega t + \delta\gamma,\qquad q=u_a/\Omega,
	\label{pha}		%72
	\end{equation}
where $\psi$ is a constant phase shift, and $\delta\gamma(t)$ is a nongrowing oscillating correction. 
Likewise, $\alpha=\alpha_0+\delta\alpha(t)$, where $\alpha_0$ is a constant to be determined, 
and $\delta\alpha(t)$ is an oscillating correction. We consider here the high-frequency signals 
of large amplitude $(\omega\tau\gg 1,\; u_a>1)$, for which both $\delta\alpha\sim 1/\omega\tau\ll 1$ 
and $\delta\gamma\sim 1/\omega\tau\ll 1$ 
can be neglected. This situation differs from the nonlocked state $(n=0)$ considered in the previous 
section, for which the contribution of $\delta\alpha$ and $\delta\gamma$ entirely determine the effect of 
the ac field on the dc $V-J$ curves. 

Eqs. (\ref{pha}) and Eqs. (\ref{aal})-(\ref{subtr}) yield the following averaged dc equations, 
	\begin{eqnarray}
	\langle\cos\gamma\rangle\sinh\alpha_0=\sqrt{h},
	\label{dca} \\		%73
	\langle\sin\gamma\rangle\sinh\alpha_0=u-n\Omega,
	\label{dcb} \\		%74
	\beta_0 = u + \langle\sin\gamma\rangle e^{-\alpha_0}.
	\label{dcc}		%75
	\end{eqnarray} 
Here the averages $\langle\sin\gamma\rangle$ and  $\langle\cos\gamma\rangle$ 
were calculated in  Appendix E:
	\begin{eqnarray}
	\langle\sin\gamma\rangle = (-1)^nJ_n(q)\sin\psi,
	\label{sin}\\		%76
	\langle\cos\gamma\rangle = (-1)^nJ_n(q)\cos\psi,
	\label{cos}		%77
	\end{eqnarray}
where $J_n(q)$ is the Bessel function. From Eqs. (\ref{dca})-(\ref{cos}), it follows that 
	\begin{eqnarray}
	\tan\psi = (u-n\Omega)/\sqrt{h}, 
	\label{psi}\\		%78
	\sinh^2\alpha_0=[h+(u-n\Omega)^2]/J_n^2(q). 
	\label{alpac}		%79
	\end{eqnarray}
Eqs. (\ref{dcc}), (\ref{psi}), and (\ref{alpac}) give the following 
dc $J-V$ characteristics at $V\approx nV_\omega$:
	\begin{equation}
	RJ=nV_\omega + (V-nV_\omega)\Bigl[1+\frac{J_n^2(q)}{h+(V-nV_\omega)^2/V_0^2}\Bigr]^{1/2}.
	\label{shap}		%80  
	\end{equation}
The behavior of $J(V)$ near the resonant voltage $V\approx nV_\omega$ is shown in Fig. 9. 
For $H\rightarrow H_0$, the $J(V)$ dependence approaches that of the   
Shapiro step for a small Josephson junction at zero field. For finite $H$, the contribution from the  
AJ vortex motion broadens the kink in $J(V)$, whose width $\Delta V = V-V_\omega$ can be estimated 
from the condition that two terms in the denominator of Eq. (\ref{shap}) become comparable.  This yields 
	\begin{equation}
	\Delta V\simeq \frac{Rc\sqrt{\phi_0H}}{8\pi\lambda^2}.
	\label{delv}		%81
	\end{equation}
The value $\Delta V$ is independent of $J_b$ and goes to zero as $T_c-T$ at $T_c$. 

The evolution of the AJ core length $L=l\alpha/\sqrt{h}$ as the dc  
voltage $V$ sweeps through the resonance at $V_\omega$ is described by Eq. (\ref{alpac}) as follows 
	\begin{equation}
	L=\frac{\l}{\sqrt{h}}\sinh^{-1}\Bigl[\frac{\sqrt{h+(u-n\Omega)^2}}{|J_n(q)|}\Bigr]
	\label{lac}		%82
	\end{equation}
The core length $L(V)$ passes through minima at the resonance voltages 
$n\hbar\omega/2e$. Notice that there are specific values of 
the ac parameter $q_n=u_a/\Omega$ which correspond to zeros of the Bessel 
function $J_n(q)$. In this case the core length diverges logarithmically, so the  
above approximation, which neglects the ac corrections $\delta\gamma\sim\delta\alpha\propto 1/\Omega$, 
becomes invalid. The blowing up of the AJ core length above $\lambda$ at $q\to q_m$ may indicate a 
conversion of the AJ into the J vortex under the action of an ac field.   
   
\section{Discussion}

In this paper solutions that describe dc and ac driven mixed  Abrikosov vortices with Josephson cores on 
high-$J_b$ grain boundaries in a magnetic field are obtained. These solutions give self-consistent distributions 
of currents circulating around moving AJ vortex structure in an exactly solvable model of the overdamped AJ 
vortex dynamics that describes both nonlinear dissipative processes in the vortex cores and magnetic interaction between 
AJ vortices along a GB. Unlike Josephson vortices whose overdamped nonlinear dynamics 
in the long junctions can be described only numerically, the dynamics of AJ vortices turned out to be 
integrable just in the overdamped limit which is most relevant to low-angle GBs.  The analytic theory of the 
ac response developed in this paper could be used to describe high-$J_b$ flux flow oscillators based 
on HTS bicrystals\cite{rf5}. 

Based on the exact AJ dynamic solutions, both the dc flux flow resistivity and the V-J characteristics 
are obtained.  The field dependence of the flux flow resistivity $R_f(H)$ shows the characteristic 
$\sqrt{H}$ behavior at low H, but then it approaches the quasiparticle limit $R$ for $H\gg H_0$ as the AJ 
cores overlap. The simplicity of Eq. (\ref{r}) for $R_f(H)$ gives a direct way of extracting the AJ core 
length $l$ and thus the intrinsic depairing current density $J_b$ and the quasiparticle 
resistance $R$ averaged over few current channels from transport measurements. Such 
measurements have indeed proven the existence of AJ vortices on $7^{\circ}$ irradiated and 
unirradiated YBCO bicrystals for which the AJ core lengths  
$\simeq 100-200 \AA$ at $55-77$K is considerably greater than $\xi(T)$, but smaller than $\lambda(T)$ \cite{let}.  
In addition, the extracted temperature dependence of $J_d(T)$ exhibited a clear 
SNS behavior $J_b=J_0(1-T/T_c)^2$, indicating a significant order parameter suppression between dislocation cores, 
even on a rather low-angle $7^{\circ}$ GB, in accordance with the model of Ref. \onlinecite{mod}.   

The fact that moving AJ vortex core can effectively probe local GB properties at the nanoscale of 
few GB dislocation spacings, makes standard transport measurements a very useful 
tool to clarify the structure of vortex core on GBs, mechanisms of current transport through GBs 
and the effect of local overdoping on $J_b$ and $R$. This method  
implies measurements of flux flow resistance $R_f(H)$ for bicrystals with the same misorientation angle 
$\vartheta$, but different local doping level. In this way, the intrinsic depairing current density 
of a GB, $J_b$  can be obtained as a function of the dopant concentration. Pinning of AJ vortices 
strongly affects the $V-J$ curve near the depinning current density, $J\approx J_{gb}$, 
but for $J\gg J_{gb}$, the differential resistivity $dV/dJ$ approaches the free flux flow 
resistivity \cite{blat}.  Thus, measuring $R_f(H)$ in the flux flow region at $J>(2-3)J_{gb}$  
enables one to avoid the analysis of multiple pinning mechanisms on real grain boundaries in HTS\cite{dcl}, 
using instead an exact flux flow theory for the interpretation of the experimental data in the region 
$J\gg J_{gb}$, where pinning is a weak perturbation. This conclusion is consistent with the experimental fact that 
that $V-J$ curves observed on HTS bicrystals are rather straight above $J_{gb}$ in a wide range of 
currents\cite{dop3,diaz,ornl,anl,albr,hogg,claus,let}. Thus, the intrinsic properties of grain boundaries can be extracted 
from the analysis of the differential resistance (but not V-J curves) measured at $J\gg J_{gb}$ using our 
solution for $R_f$ which neglects pinning. A similar approach was used to measure the flux flow 
resistivity of pinned A vortices driven by strong current pulses well above $J_c$\cite{kunchur}.  
Since $J_{gb}$ is by 2-3 orders of magnitudes below the intrinsic $J_b$, the pinning region is 
much smaller than the scale of Fig. 2, so the linear $R_f$ can be used to fit the data.   

Measuring the local ac response of a GB can bring additional advantages over the dc 
measurements in which the dc current acts both on AJ on GB and A vortices in the grains. 
Indeed, using a scanning localized microwave source\cite{rf2}, it would be possible to 
probe the ac dynamics of AJ on GB, minimizing the effect of pinned A vortices. For instance, 
the AJ core size and thus the local $J_b$ can be extracted from ac measurements of 
higher voltage harmonics, as described in the previous section.   

The behavior of AJ vortices in a strong dc field $H\gg H_{c1}$ considered in this paper  
is most relevant for bulk HTS. However, of much interest for superconducting electronics
is also the electromagnetic response of HTS polycrystaline films in comparatively weak rf fields 
for which vortices mostly penetrate the network of GBs.  In equilibrium, this regime corresponds to the field range  
$H_{c1b}<H<H_{c1}$, where $H_{c1}=(\phi_0/4\pi\lambda^2)(1-N)[\ln(\lambda/\xi) + 0.5]$ is the lower critical field for 
intragrain A vortices \cite{abr}, and $H_{c1b}=(\phi_0/4\pi\lambda^2)(1-N)[\ln(\lambda/l)+0.423]$ is the lower 
critical field of AJ vortices\cite{ag}, and N is the demagnetizing factor. For a film in a perpendicular field, 
the AJ vortex spacing $a=(\phi_0/B)^{1/2}$ can now significantly vary along GB, since the normal component of the 
local magnetic induction $B(x,y)$ is now determined by highly nonuniform distribution of the Meissner surface  
currents and a dome-like vortex distribution due to the geometrical barrier\cite{geomb1,geomb2}. In this case the 
results of this paper based on the solution (\ref{theta}) with constant $k=2\pi/a(B)$ may be used if $B(x)$ varies 
slowly over the inter vortex spacing $a(x)$, thus the parameter $h$ in the above formulas should be 
replaced by its local value $h(x)=B(x)/H_0$.

\section{Acknowledgments}
This work was supported by the NSF  MRSEC (DMR 9214707),
AFOSR MURI (F49620-01-1-0464). 

\appendix 

\section{Derivation of Eq. (\ref{nj})}

To calculate the 2D current distribution around a planar Josephson contact in the xz plane at $y=0$, 
it is convenient to use the scalar stream function $\psi(x,y)$ so that $J_x=\partial_y\psi$, 
$J_y=-\partial_x\psi$, whence  
	\begin{eqnarray}
	\frac{\partial\psi}{\partial y}=\frac{c}{4\pi\lambda^2}
	\left(\frac{\phi_0}{2\pi}\frac{\partial\varphi}{\partial x}- A_x\right), 
	\label{a1} \\
	\frac{\partial\psi}{\partial x}=-\frac{c}{4\pi\lambda^2}
	\left(\frac{\phi_0}{2\pi}\frac{\partial\varphi}{\partial y}- A_y\right), 
	\label{a2}
	\end{eqnarray}
where ${\bf A}$ is the vector potential, and $\varphi$ is the phase of the 
order parameter. From Eq. (\ref{a1}), it follows that the parallel 
component $J_x(x,+0)-J_x(x,-0)=(c\phi_0/8\pi^2\lambda^2)\theta '$ is 
discontinuous for any nonuniform distribution of the phase difference $\theta(x)=\varphi(x,+0)-\varphi(x,-0)$ 
along GB. This results in the following boundary condition for the stream function at a GB:
	\begin{equation}
	\partial_y\psi(x,+0)-\partial_y\psi(x,-0)=
	(c\phi_0/8\pi^2\lambda^2)\theta '.
	\label{a3}
	\end{equation}
Excluding $\varphi$ from Eqs. (\ref{a1}) and (\ref{a2}), we obtain
	\begin{equation}
	\nabla^2\psi-\frac{cH}{4\pi\lambda^2}= 
	\frac{c\phi_0\theta'}{8\pi^2\lambda^2}\delta(y),
	\label{a4}
	\end{equation}
where the delta function $\delta(y)$ in the right-hand side provides the boundary 
condition (\ref{a3}), and $H=\nabla_z\times{\bf A}$ is the z-component of the magnetic field, 
related to the stream function $\psi$ by the Biot-Savart law:
	\begin{equation}
	H({\bf r})=\frac{1}{c}
	\int_{V}\frac{[(x-x')\partial_{x'}\psi+(y-y')\partial_{y'}\psi]d^3{\bf r'}}{[(x-x')^2+(y-y')^2+(z-z')^2]^{3/2}}.
	\label{a5}
	\end{equation}
Eqs. (\ref{a4}) and (\ref{a5}) give an integro-differential equation for $\psi(x,y)$ 
which can be solved by the Fourier transform. 

For an infinite (along ${\bf H}$) sample, the local relation 
$\psi(x,y) = cH(x,y)/4\pi$ holds, thus Eq. (\ref{a4}) becomes 
the London equation (\ref{london}) for $H(x,y)$. For a thin film of thickness 
$d\ll\lambda$, one can put $z=z'=0$ in Eq. (\ref{a5}), then the integration over $z'$ 
gives the factor $d$, and Eq. (\ref{a5}) gives the following relation between the 
Fourier components of $H$ and $\psi$:
	\begin{equation}
	H_q = (2\pi qd/c)\psi_q,
	\label{a7}
	\end{equation}
where $q^2=q_x^2+q_y^2$. The Fourier transform of Eq. (\ref{a4}) yields
	\begin{equation}
	\psi_q=-\frac{c\phi_0\theta^{\prime}(q_x)}{8\pi^2\lambda^2(q^2+q/\tilde{\lambda})}
	\label{a8}
	\end{equation}
where $\tilde{\lambda}=2\lambda^2/d$. Making the inverse Fourier transform of Eq. (\ref{a8}) 
and using the boundary condition $\psi^{\prime}(x,0)=J_b\sin\theta + \hbar\dot\theta/2eR$, 
we arrive at the nonlocal Eq. (\ref{th}) for $\theta(x,t)$ in which $K_0(|x-u|/\lambda)$ should 
be replaced by the kernel $\tilde{K}(|x-u|/\tilde{\lambda})$, where\cite{ms,ames}
	\begin{equation}
	\tilde{K}(s)=\int_0^{\infty}\frac{e^{-sp}dp}{\sqrt{1+p^2}}
	\label{a9}
	\end{equation}

Despite different behaviors of $K_0(|x-u|/\lambda)$ and $\tilde{K}(|x-u|/\tilde{\lambda})$, 
at large distances, they both have the same logarithmic singularity at $x=u$, because   
at short distances $|u-x|\ll\lambda$, screening is inessential so the vector potential ${\bf A}$ 
in Eqs. (\ref{a1})-(\ref{a2}) and the field $H$ in Eq. (\ref{a4}) can be neglected. 
Then Eq. (\ref{a4}) becomes a 2D Poisson equation whose solution is   
	\begin{equation}
	\psi(x,y)= \frac{c\phi_0}{32\pi^3\lambda^2}\int_{-\infty}^\infty\theta^{\prime}(u)\ln[y^2+(x-u)^2]du
	\label{a10}
	\end{equation}
both for thin films and bulk samples. 

\section{Exact solution}

To show that Eq. (\ref{theta}) is indeed an exact solution of Eq. (\ref{nj}), 
we substitute Eq. (\ref{tprime})-(\ref{sint}) into Eq. (\ref{nj}) in which the integral 
is evaluated using the Hilbert transform\cite{seeger,alfimov}
	\begin{equation}
	\int_{-\infty}^{\infty}\frac{\sinh\alpha dy}{(x-y)(\cosh\alpha-\cos ky)}
	=\frac{\pi\sin kx}{\cosh\alpha-\cos kx}
	\label{b1}
	\end{equation}
Eq. (\ref{nj}) then reduces to the algebraic form $C_1\cos\zeta+C_2\sin\zeta+C_3=0$, where 
the coefficients $C_i$ depend only on time via $\alpha(t)$, $\gamma(t)$, and ${\dot x}_0(t)$. 
Equating all $C_i$ to zero, we arrive at Eqs. (\ref{al})-({\ref{u}}). 

Useful relations for the steady-state vortex propagation 
can be obtained from Eqs. (\ref{al})-({\ref{u}}) with ${\dot\gamma}={\dot\alpha}=0$:
	\begin{eqnarray}
	\sinh\alpha\cos\gamma=\sqrt{h}, 
	\label{b2}\\
	\sin\gamma\cosh\alpha=\beta_0(t),
	\label{b3}\\ 
	s\sqrt{h}=-\sin\gamma\sinh\alpha.
	\label{b4}
	\end{eqnarray} 
Adding squared Eqs. (\ref{b2}) and (\ref{b4}) yields 
	\begin{equation}
	\sinh^2\alpha=(1+s^2)h.
	\label{b5}
	\end{equation}
Subtracting squared Eqs. (\ref{b3}) and (\ref{b4}) yields
	\begin{equation}
	\sin^2\gamma=\beta_0^2-s^2h,
	\label{b6}
	\end{equation}
Substituting Eqs. (\ref{b5}) and (\ref{b6}) back to 
Eq. (\ref{b4}) gives the following equation for $s$ 
	\begin{equation}
	\beta_0^2=hs^2+\frac{s^2}{1+s^2}.
	\label{b7}
	\end{equation}
The use of the relation $s\sqrt{h}=V/J_bR$ reduces Eq. (\ref{b7}) to Eq. (\ref{jv}).    
Substituting Eq. (\ref{b7}) into Eq. (\ref{b6}) yields 
	\begin{equation}
	\sin\gamma=\frac{s}{\sqrt{1+s^2}},\qquad \cos\gamma=\frac{1}{\sqrt{1+s^2}}	
	\label{b8}
	\end{equation} 

These results can also be obtained from the complex representation of the 
dc equations (\ref{comp})
	\begin{equation}
	\sin z_0=\beta_0+i\sqrt{h}.
	\label{b9}
	\end{equation}
The vortex velocity is given by the second Eq. (\ref{comp}) 
	\begin{equation}
	s\sqrt{h}=\mbox{Im}[1-(\beta_0+i\sqrt{h})]^{1/2},
	\label{b10}
	\end{equation}
from which Eq. (\ref{sv}) readily follows.

\section{$R_f$ for J vortices}

The sine-Gordon equation for a periodic J vortex structure moving with a 
constant velocity $v$ has the form
	\begin{equation}
	\lambda_J^2\partial_{\zeta\zeta}\theta+v\tau\partial_\zeta\theta-\sin\theta+\beta=0,
	\label{c1}
	\end{equation}
where the $\zeta=x-vt$.
Multiplying Eq. (\ref{c1}) by $\partial_\zeta\theta$ and then integrating from $\zeta=0$ to $\zeta=a$, using the flux 
quantization condition $\theta(\zeta+a)-\theta(\zeta)=2\pi$, and the periodicity condition 
$\partial_\zeta\theta(\zeta+a)=\partial_\zeta\theta(\zeta)$ for $H(\zeta)$, we arrive at the 
following equation for $v$:
	\begin{equation}
	2\pi\beta=-v\tau\int_0^a(\partial_\zeta\theta)^2d\zeta
	\label{c2}
	\end{equation}
Expressing the vortex velocity $v(J)$ in Eq. (\ref{c2}) via the dc voltage $V$ from Eq. (\ref{vol})
yields the flux flow resistivity $R_f=V/J$ in the form
	\begin{equation}
	R_f=4\pi^2R/\Bigl[a\int_0^a(\partial_\zeta\theta)^2d\zeta\Bigr]
	\label{c3}
	\end{equation}
For the linear flux flow state at $\beta\ll 1$, 
$\theta(\zeta)$ in Eq. (\ref{c3}) can be replaced with  
static $\theta_0(x)$ for which  
	\begin{equation}
	\lambda_J^2\theta_0''-\sin\theta_0=0,
	\label{c4}
	\end{equation}
The first integral of Eq. (\ref{c4}) has the form
	\begin{equation}
	(\lambda_J\theta_0')^2=4[p^{-2}-\cos^2\theta_0/2]
	\label{c5}
	\end{equation}
The periodic solutions of Eq. (\ref{c5}) can be expressed in terms of the Jacobi elliptic function, 
$\theta_0(x)=\pi+2am(x/p\lambda_J)$, where $0<p<1$ is a parameter which is related to 
the applied magnetic field $H$ and the magnetic induction $B$ by Eqs. (\ref{bhj})\cite{os}. 
The period $a$ of the J structure is given by 
	\begin{equation}
	a=2\lambda_JpK(p), \qquad 2\lambda aB=\phi_0,
	\label{c6}
	\end{equation}
where $K(p)$ and $E(p)$ are the complete elliptic integrals\cite{abr},
	\begin{eqnarray}
	K(p)=\int_0^{\pi/2}(1-p^2\sin^2\theta)^{-1/2}d\theta,
	\label{c7}\\ 
	E(p)=\int_0^{\pi/2}(1-p^2\sin\theta^2)^{1/2}d\theta
	\label{c8}
	\end{eqnarray}
Substituting Eq. (\ref{c5}) into Eq. (\ref{c3}) and changing integration 
$\int_0^adx\to \int_0^{2\pi}d\theta/\theta'$, we obtain 
	\begin{equation}
	R_f=\frac{\pi^2Rp\lambda_J}{2aE(p)}
	\label{c9}
	\end{equation}
Eqs. (\ref{c6}), (\ref{c9}) and (\ref{bhj}) give Eq. (\ref{rfj}).

\section{Linear response}

Linearized Eqs. (\ref{al})-(\ref{u}) have the form 
	\begin{eqnarray}
	\tau\delta\dot\alpha+\delta\alpha\cosh\alpha\cos\gamma-\delta\gamma\sinh\alpha\sin\gamma=0,
	\label{d1} \\
	\tau\delta\dot\gamma+\delta\gamma\cos\gamma\cosh\alpha+\delta\alpha\sin\gamma\sinh\alpha=
	\beta_\omega e^{i\omega t},
	\label{d2} \\
	k\tau\delta v=-\delta\gamma\cos\gamma\sinh\alpha-\delta\alpha\sin\gamma\cosh\alpha
	\label{d3}
	\end{eqnarray}
These equations give the following Fourier components of $\delta\alpha$, $\delta\gamma$, 
and $\delta s$:
	\begin{eqnarray}
	\delta\alpha_\omega = \frac{\sinh\alpha\sin\gamma\delta\beta_\omega}
	{(i\Omega +\cosh\alpha\cos\gamma)^2+\sinh^2\alpha\sin^2\gamma},
	\label{d4} \\
	\delta\gamma_\omega = \frac{(i\Omega+\cosh\alpha\cos\gamma)\delta\beta_\omega}
	{(i\Omega +\cosh\alpha\cos\gamma)^2+\sinh^2\alpha\sin^2\gamma},
	\label{d5} \\
	\sqrt{h}\delta s_\omega = -\frac{\sinh\alpha(i\Omega\cos\gamma+\cosh\alpha)\delta\beta_\omega}
	{(i\Omega +\cosh\alpha\cos\gamma)^2+\sinh^2\alpha\sin^2\gamma},
	\label{d6}
	\end{eqnarray} 
where $\Omega = \omega\tau$. Substituting $\delta s$ and $\delta\gamma$ into Eq. (\ref{vol}) gives
the complex ac resistance $R_\omega = V_\omega/J_a$:
	\begin{equation}
	\frac{R_\omega}{R}=\frac{\sinh\alpha\cosh\alpha -\Omega^2+i\Omega e^{\alpha}\cos\gamma}
	{\sinh^2\alpha+\cos^2\gamma-\Omega^2+2i\Omega\cosh\alpha\cos\gamma}
	\label{d7}
	\end{equation}    	
Eq. (\ref{d7}) reduces to Eq. (\ref{rdin}), using Eqs. (\ref{b5})-(\ref{b8}).

\section{Separation of fast and slow variables}

The time averages of 
	\begin{eqnarray}
	\langle\sin\gamma\rangle=\langle(\sin\psi\cos n\omega t+\cos\psi\sin n\omega t)\cos(q\sin\omega t) 
	\nonumber \\
	+(\cos\psi\cos n\omega t-\sin\psi\sin n\omega t)\sin(q\sin\omega t)\rangle
	\label{e1}
	\end{eqnarray}
	\begin{eqnarray}
	\langle\cos\gamma\rangle=\langle(\cos\psi\cos n\omega t-\sin\psi\sin n\omega t)\cos(q\sin\omega t) 
	\nonumber \\
	-(\sin\psi\cos n\omega t+\cos\psi\sin n\omega t)\sin(q\sin\omega t)\rangle,
	\label{e2}
	\end{eqnarray}
can be calculated, using the identities \cite{abr}
	\begin{eqnarray}
	\cos(q\sin\omega t)=J_0(q)+2\sum_{k=1}^{\infty}J_{2k}(q)\cos(2k\omega t),
	\label{e3}\\ 
	\sin(q\sin\omega t)=2\sum_{k=0}^{\infty}J_{2k+1}(q)\sin[(2k+1)\omega t],
	\label{e4}
	\end{eqnarray}
where $J_k(q)$ is the Bessel function.  Averaging Eqs. (\ref{e1}) and (\ref{e2}) yields Eqs. (\ref{sin}) and (\ref{cos})  
because only one resonant term with $2k=n$ or $2k+1=n$ in the sums (\ref{e3}) 
and (\ref{e4}) gives a nonzero dc contribution to $\langle\sin\gamma\rangle$ and $\langle\cos\gamma\rangle$.

\newpage

\begin{figure}				%FIG1
%\epsfxsize= 0.8\hsize  
%\centerline{
%\vbox{
%\epsffile{fig1.eps} 
%}}
%\vskip \baselineskip
\caption{Current streamlines around AJ vortices on a GB (dashed line) and 
the bulk A vortices in the grains, calculated from Eq. (\ref{he}) for $l=0.2a$.
}
\label{fig.1}
\end{figure}

\begin{figure}				%FIG2
%\epsfxsize= 0.8\hsize  
%\centerline{
%\vbox{
%\epsffile{fig2.eps} 
%}}
%\vskip \baselineskip
\caption{The $V-J$ curves calculated from Eq. (\ref{vj}) for different magnetic fields $h=H/H_0$: 
$0.01$(1), $0.05$(2), $0.1$(3), $0.5$(4), $10$(5). 
Inset shows the field dependence of the flux flow resistance $R_f(B)$ given by Eq. (\ref{r}).
}
\label{fig.2}
\end{figure}

\begin{figure}				%FIG3
%\epsfxsize= 0.7\hsize  
%\centerline{
%\vbox{
%\epsffile{fig3.eps} 
%}}
%\vskip \baselineskip
\caption{Dependencies of $\Omega_0$ and $\eta$ on the normalized dc current 
density $\beta_0=J/J_b$ for $H=0.05H_0$.
}
\label{fig.3}
\end{figure}

\begin{figure}				%FIG4
%\epsfxsize= 0.7\hsize  
%\centerline{
%\vbox{
%\epsffile{fig4.eps} 
%}}
%\vskip \baselineskip
\caption{Field dependence of the ac dissipated power $Q(H)$ for 
different dimensionless frequencies $\Omega=\omega\tau$.
}
\label{fig.4}
\end{figure}

\begin{figure}				%FIG5
%\epsfxsize= 0.7\hsize  
%\centerline{
%\vbox{
%\epsffile{fig5.eps} 
%}}
%\vskip \baselineskip
\caption{Flux flow resonance line in Im$R_\omega(\Omega)$ calculated from Eq. (\ref{rdin}) 
for $\beta=4$ and different magnetic fields $H/H_0$: 0.05, 0.4, and 2 (from bottom to top curve, respectively).
}
\label{fig.5}
\end{figure}

\begin{figure}				%FIG6
%\epsfxsize= 0.7\hsize  
%\centerline{
%\vbox{
%\epsffile{fig6.eps}
%}}
%\vskip \baselineskip
\caption{Manifestation of the flux flow resonance on the averaged dc $J-V$ characteristic for 
$\omega\tau=0.1$, $H=0.01H_0$, and different ac amplitudes, $(J_a/2J_b)^2$: 0 (1); 0.05 (2); 0.1 (3).
}
\label{fig.6}
\end{figure}

\begin{figure}				%FIG7
%\epsfxsize= 0.7\hsize  
%\centerline{
%\vbox{
%\epsffile{fig7.eps} 
%}}
%\vskip \baselineskip
\caption{The power $Q$ dissipated on a GB as a function of the ac amplitude, where   
$Q_a=2J_b^2R_f/\pi$.
}
\label{fig.7}
\end{figure}

\begin{figure}				%FIG8
%\epsfxsize= 0.7\hsize  
%\centerline{
%\vbox{
%\epsffile{fig8.eps} 
%}}
%\vskip \baselineskip
\caption{The field dependencies of the amplitude of the third harmonics $V_3(H)$ 
normalized to $V_n=RJ_a^3/16J_b^2$ and the flux flow resistance $R_f(H)$.
}
\label{fig.8}
\end{figure}

\begin{figure}
%\epsfxsize= 0.7\hsize  
%\centerline{				%FIG9
%\vbox{
%\epsffile{fig9.eps} 
%}}
%\vskip \baselineskip
\caption{Quasi-step on the averaged $J-V$ characteristic at $V\approx V_\omega$
for $q=3$, $n=1$, $V_{\omega}/RJ_b=1$, and different magnetic fields $H/H_0$: 0 (1),  0.01 (2), and 0.1 (3).
}
\label{fig.9}
\end{figure}

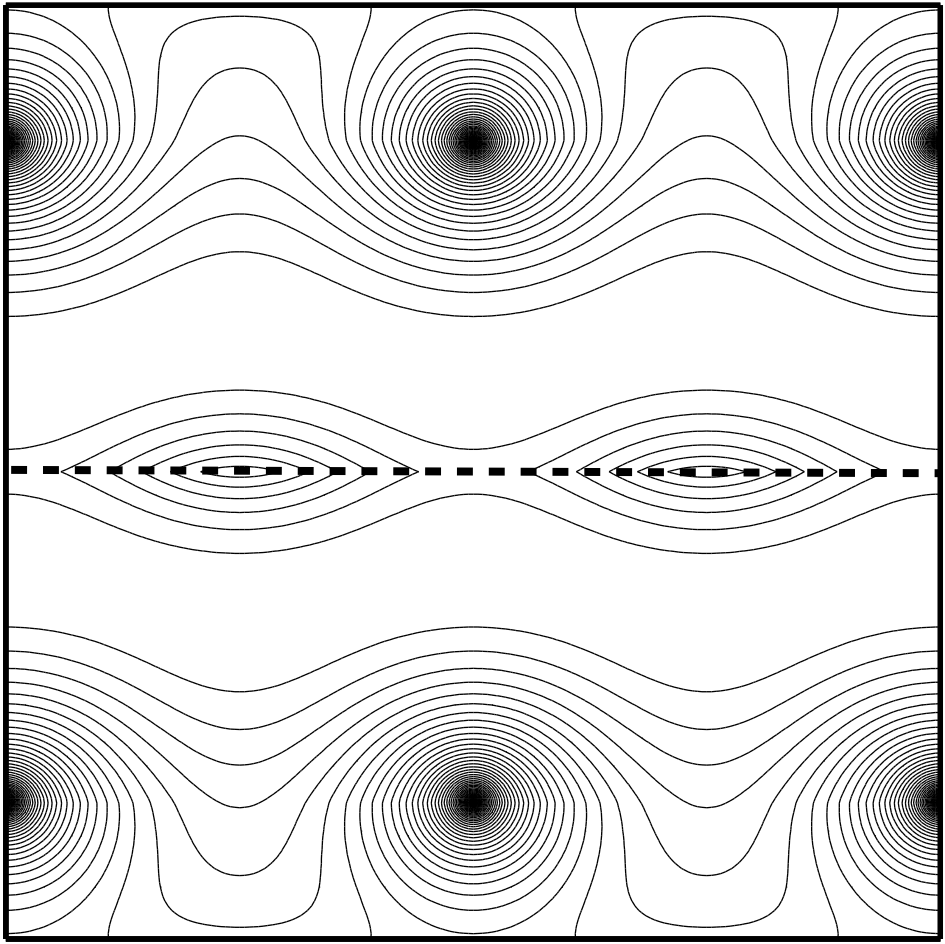
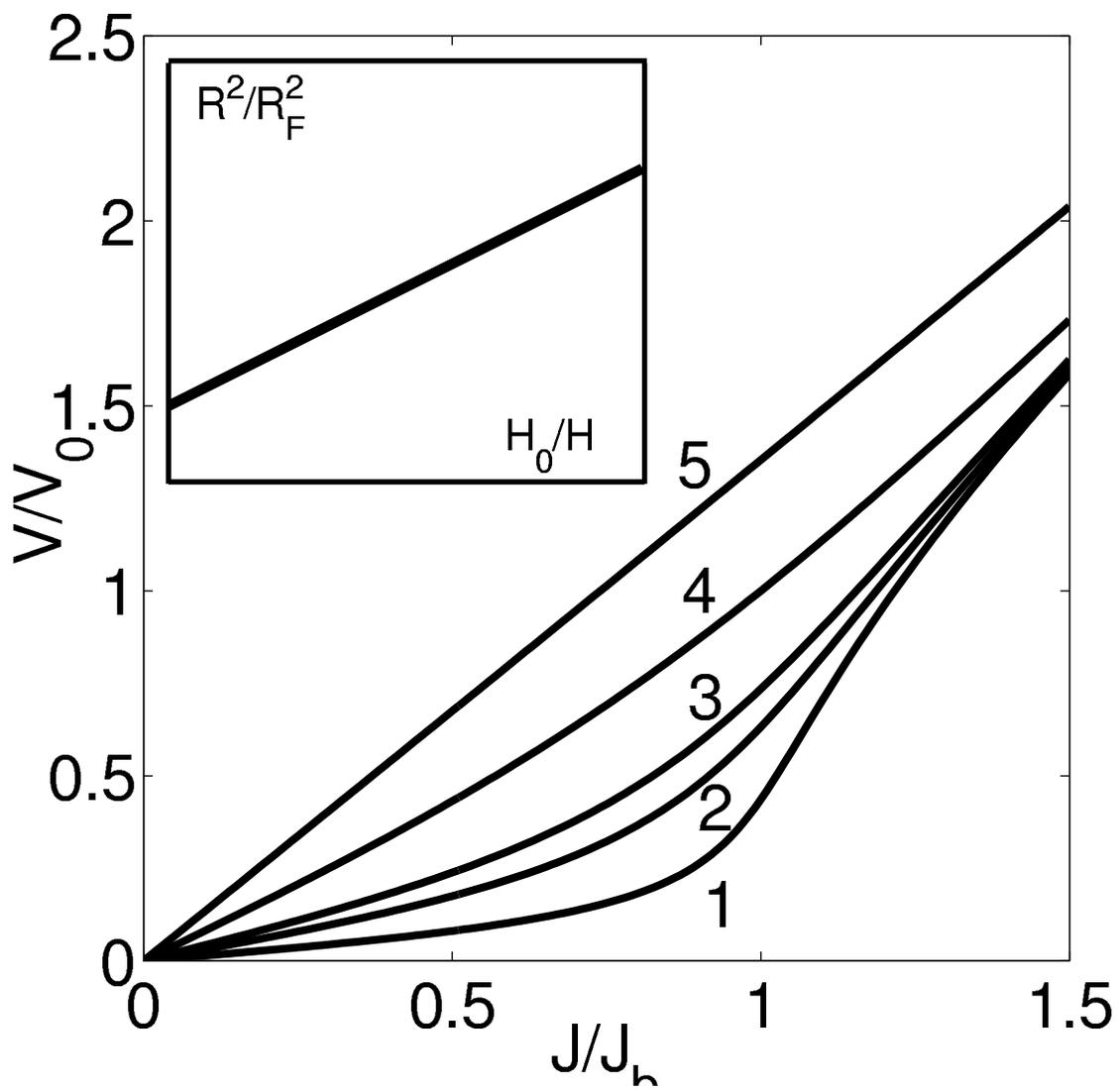
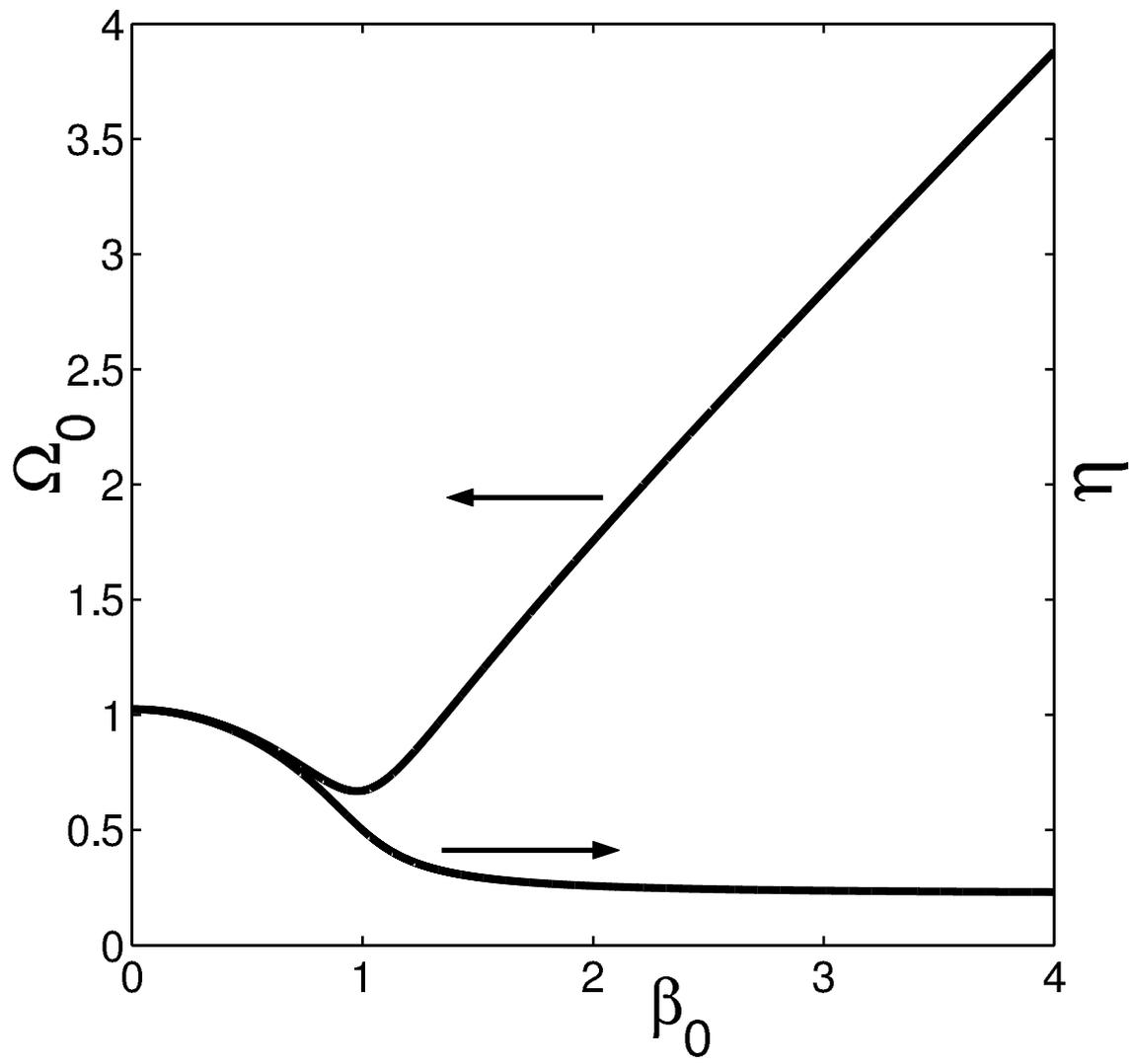
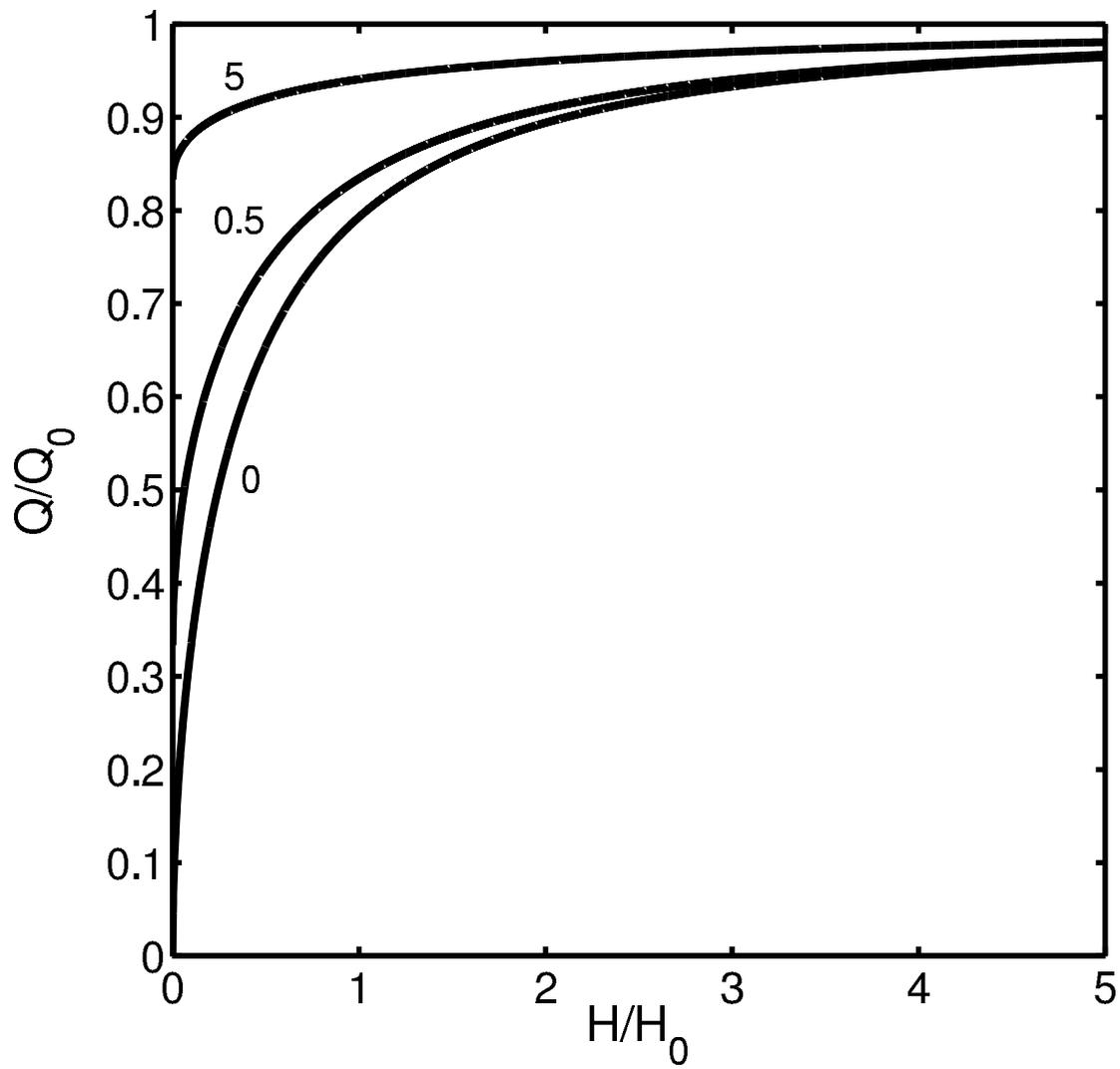
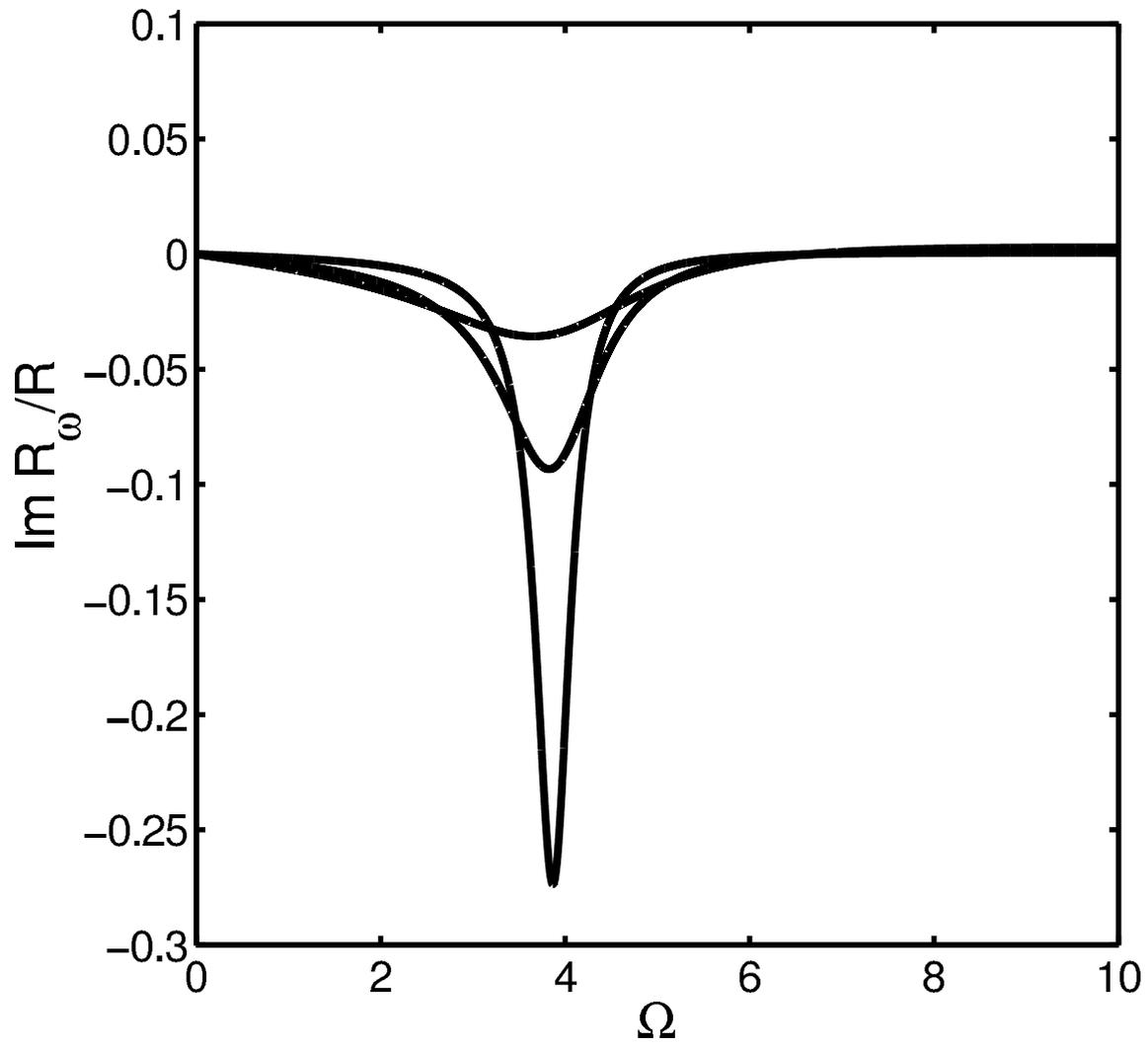
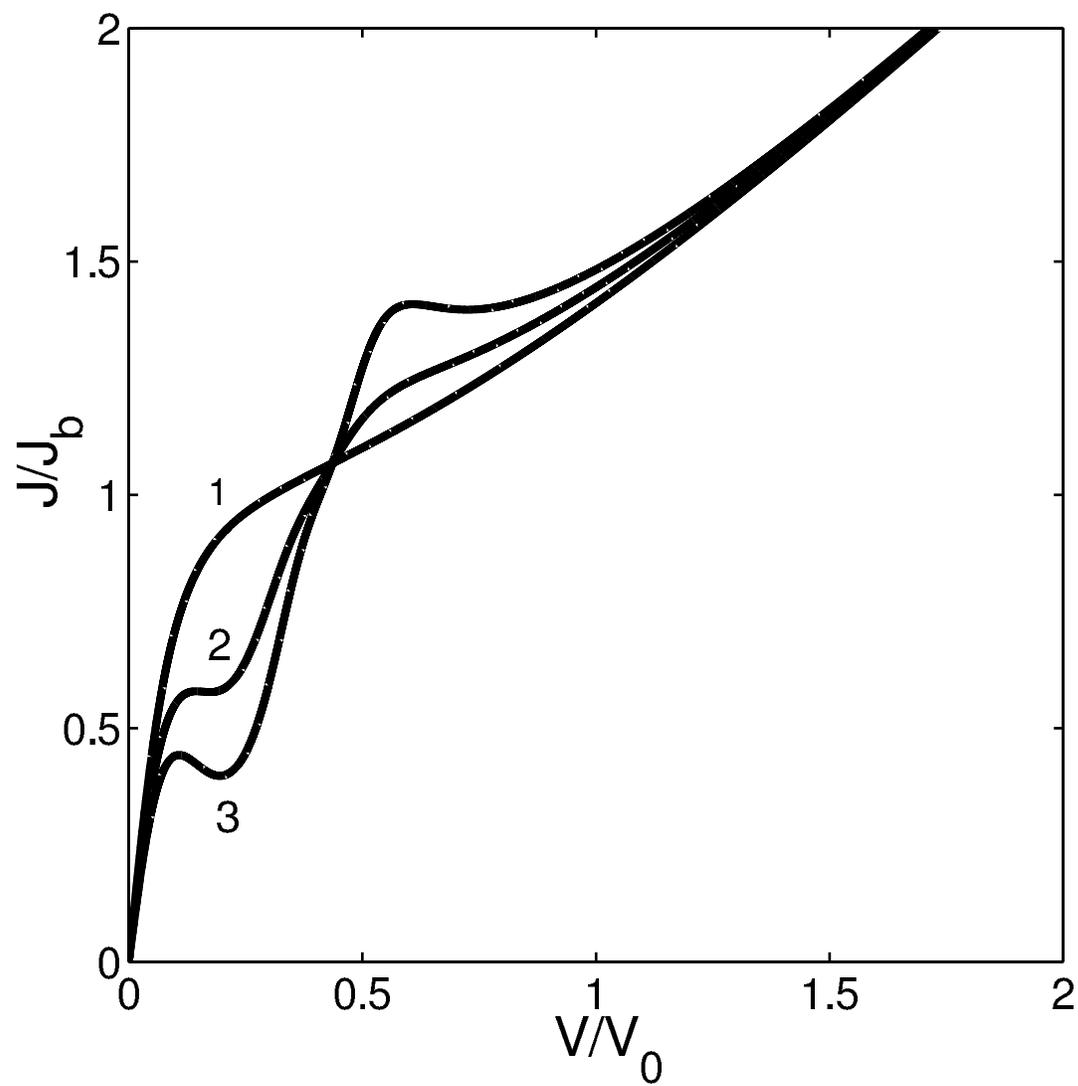
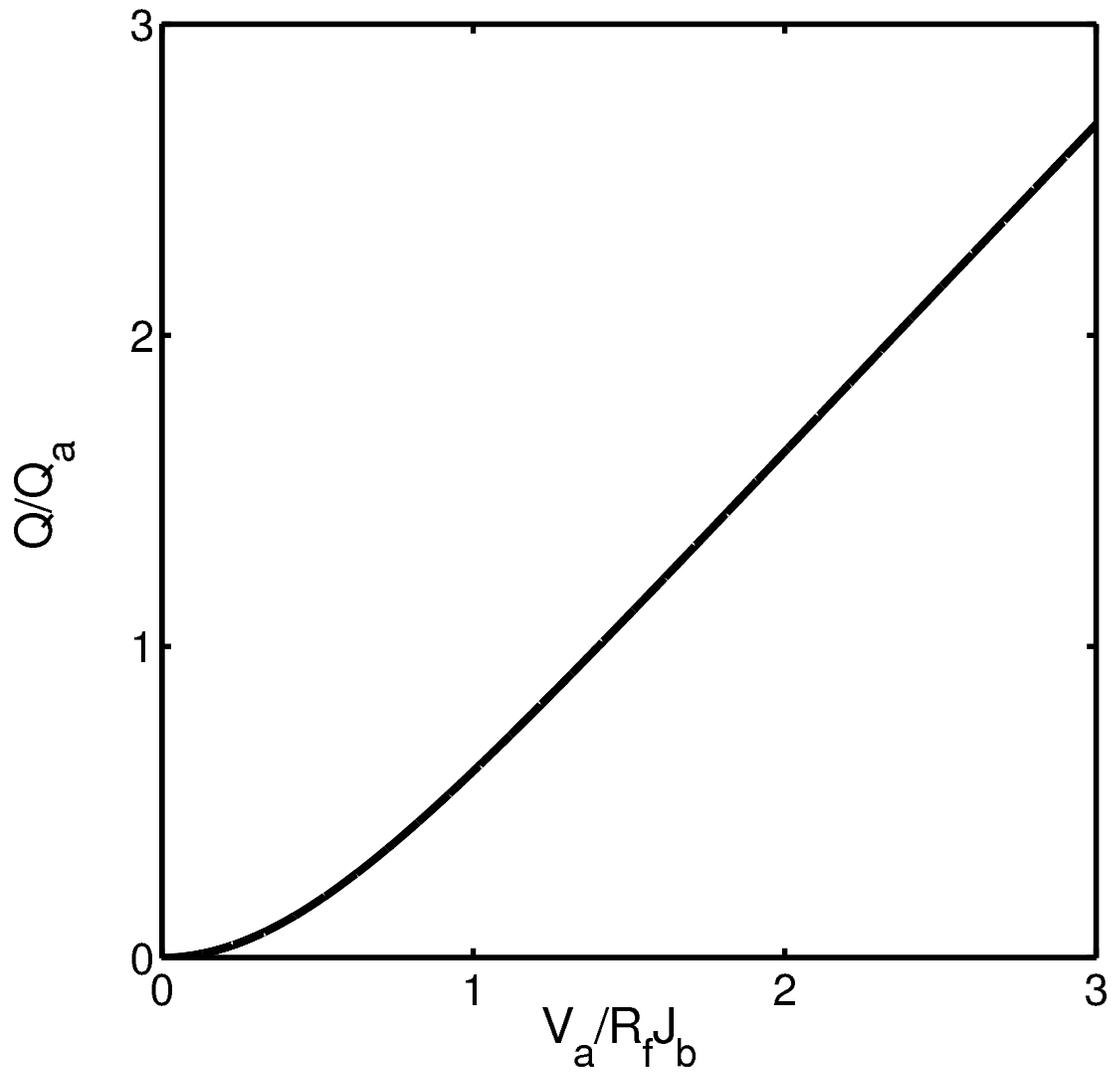
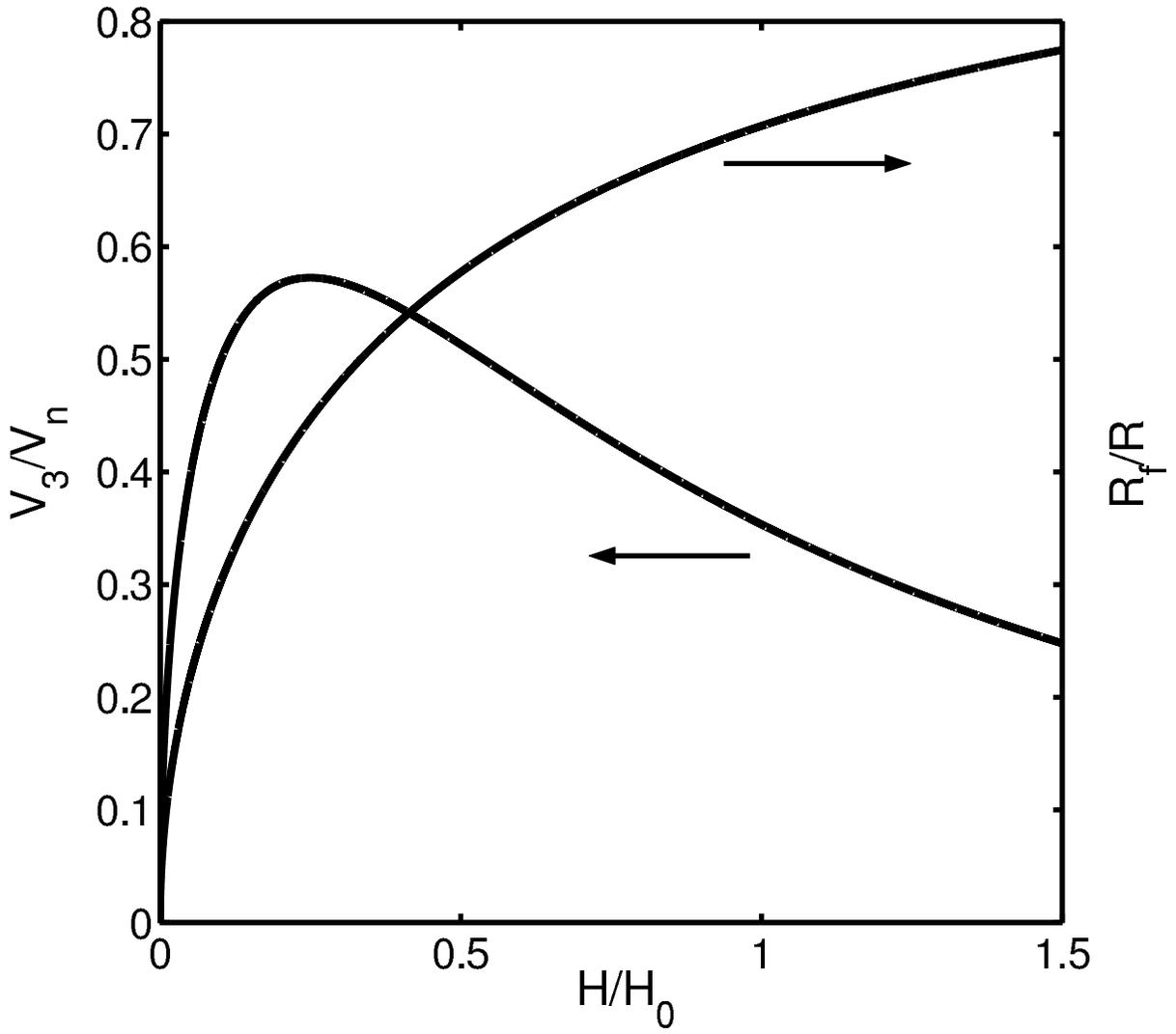
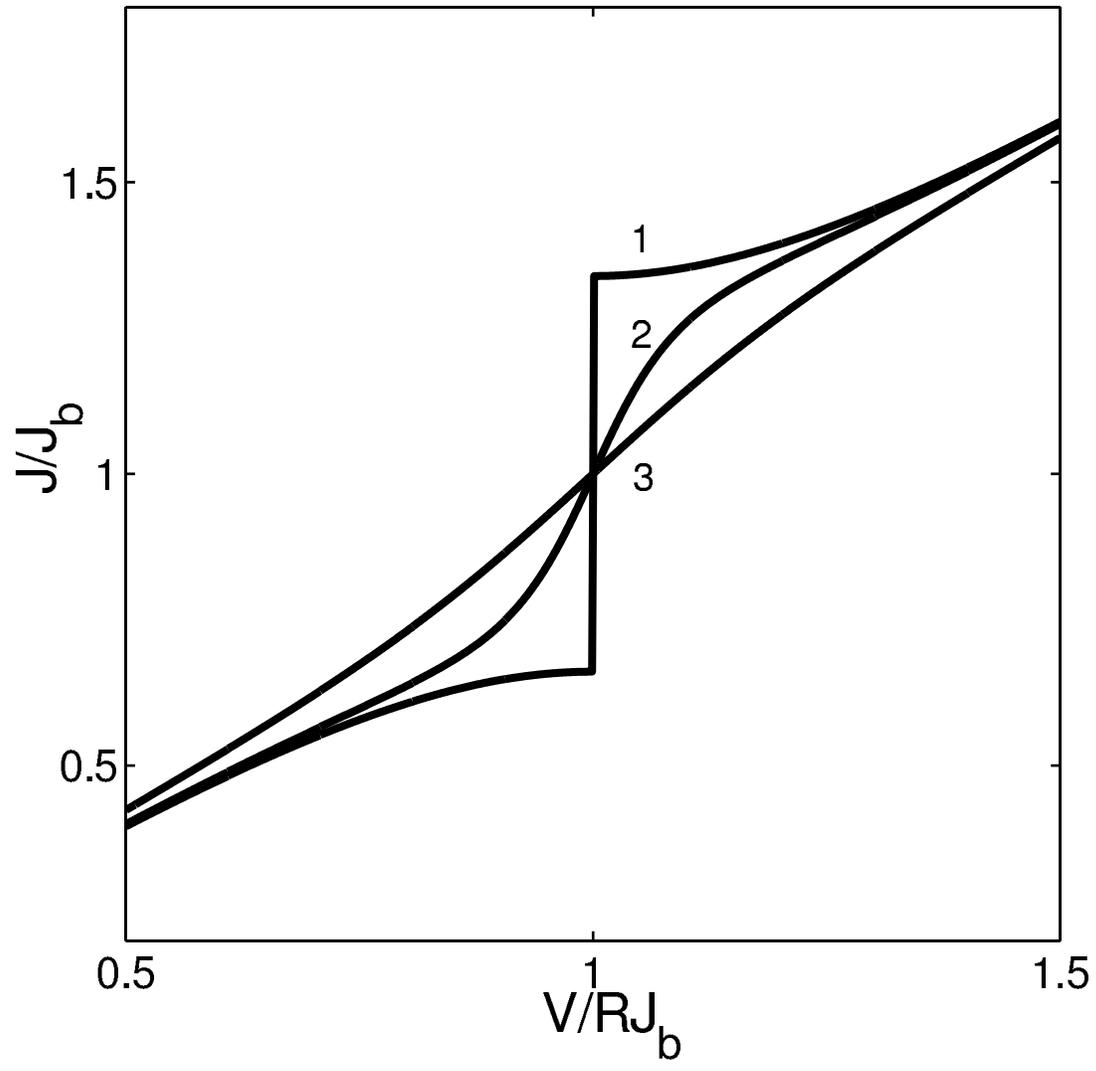
\end{document}